\documentclass[dvips,12pt]{article}
\usepackage{amsmath,amssymb,amsfonts,amscd,theorem,xspace,pstricks,pst-node,pst-coil}
\usepackage[latin1]{inputenc}
\def\simarrow{\mathrel{\raise -0.5mm\hbox{$\sim$}}\hspace{-1.8mm}{\rightarrow} } 

\def\bsimarrow{\leftarrow\hspace{-0.7mm}\mathrel{\raise -0.5mm\hbox{$\backsim$}} }

\def\bt{\begin{tabular}}
\def\te{\end{tabular}}

\def\lettrine#1#2#3{\noindent\hangindent#1\hangafter-#2
\hskip-#1\smash{\hbox to #1{#3\hfill}}\ignorespaces}

\newcommand{\To}[1]{\mathop{\to}\limits_{#1}}

\def\BM{\begin{pmatrix}}
\def\EM{\end{pmatrix}}

\def\pt{\mathrel{\scriptstyle \bullet}}

\def\d=f{\buildrel\hbox{\scriptsize d\'{e}f}\over \Longleftrightarrow}

\def\cit{\text{\it I\hskip -6ptC\/}}

\def\square{\hfill\hbox{\vrule height .9ex width .8ex depth -.1ex}}
\def\rit{\text{\it I\hskip -2pt  R}}

\def\rl {\rit^{\hskip 1pt\ell}}

\def\Bd{{\text B}}

\def\Ds{{\cal D}}
\def\Ed{{\text E}}

\def\be{\begin{equation}}
\def\ee{\end{equation}}
\def\beqn{\begin{eqnarray}}
\def\eeqn{\end{eqnarray}}
\def\nobeqn{\begin{eqnarray*}}
\def\noeeqn{\end{eqnarray*}}
\def\ba{\left(\begin{array}}
\def\ea{\end{array} \right) }

\def\bpr{\paragraph{Proof.}}

\def\epr{\square\vskip 6pt}
\def\eop{\hbox{\vrule height .9ex width .8ex depth -.1ex}}

\def\o{\overline}
\def\and{\; \mbox{and} \;}

\newcommand{\half}{\frac{1}{2}}

\def\hfl#1#2{\smash{\mathop{\hbox to 12mm{\rightarrowfill}}
\limits^{\scriptstyle #1}_{\scriptstyle #2}}}

\def\Be{\begin{enumerate}}
\def\Ee{\end{enumerate}}

\def\Bena{\begin{enumerate}
\def\labelenumi{\theenumi)}
\def\theenumi{\arabic{enumi}}
\def\labelenumii{\theenumii)}
\def\theenumii{\alph{enumii}}}

\def\Bean{\begin{enumerate}
\def\labelenumii{\theenumii)}
\def\theenumii{\arabic{enumii}}
\def\labelenumi{\theenumi)}
\def\theenumi{\alph{enumi}}}

\def\Bero{\begin{enumerate}
\def\labelenumii{\theenumii)}
\def\theenumii{\arabic{enumii}}
\def\labelenumi{(\theenumi)}
\def\theenumi{\roman{enumi}}}

\def\BeRo{\begin{enumerate}
\def\labelenumii{\theenumii)}
\def\theenumii{\arabic{enumii}}
\def\labelenumi{(\theenumi)}
\def\theenumi{\Roman{enumi}}}

\def\Bi{\vskip 11pt\begin{itemize}\itemsep=18pt}
\def\bi{\begin{itemize}}
\def\Ei{\end{itemize}\vskip 11pt}
\def\ei{\end{itemize}}

\def\Bd{\begin{description}}
\def\Ed{\end{description}}

\def\R{\right}
\def\L{\left}
\def\F{\frac}

\def\prod{\mathop{\Pi}\limits}
\def\sum{\mathop{\Sigma}\limits}
\def\bigoplus{\mathop{\oplus}\limits}

\def\Boxplus{\mathop{\boxplus}\limits}

\def\mfg#1{``~$#1$~''\xspace}

\def\vec{\overrightarrow}

\newcommand{\myendofpart}{\ifodd\value{page}\vfill\eject
\thispagestyle{empty}
\fi}

\def\Bi{\begin{itemize}}

\def\FF{{\mathbb{F}\,}}
\def\cit{{\mathbb{C}\,}}
\def\rit{{\mathbb{R}\,}}
\def\NN{{\mathbb{N}\,}}

\def\labelenumi{\alph{enumi})}
\def\theenumi{\alph{enumi}}
\def\labelenumii{\arabic{enumii}.}
\def\theenumii{\arabic{enumii}}

\def\BeAn{\begin{enumerate}
\def\labelenumii{\theenumii)}
\def\theenumii{\arabic{enumii}}
\def\labelenumi{\theenumi)}
\def\theenumi{\Alph{enumi}}}

\def\Repsp{\operatorname{Repsp}}

\def\resp#1{(resp. #1~)}
\def\rresp#1{\qquad \mbox{(resp.} \quad #1\ )}

\def\RL{_{R\times L}}

\def\FRepsp{\operatorname{FRepsp}}
\def\GL{\operatorname{GL}}

\def\Diff{\operatorname{Diff}}

\def\MM{{\mathbb{M}\,}}

\def\EE{E}

{\theorembodyfont{\rmfamily}
  }
{\theorembodyfont{\rmfamily}
  }

{\theorembodyfont{\rmfamily}
  }
{\theorembodyfont{\rmfamily}
  }

{\theorembodyfont{\rmfamily}
  }

\def\lr{left (resp. right) }
\def\rl{right (resp. left) }
\def\gl{{\mathfrak{g}}\ell}

\def\bbf{\boldmath\bf}

\def\To{\begin{CD} @>>>\end{CD}}

\begin{document}

\pagestyle{myheadings}

\def\thepage{\arabic{page}}

{\pagestyle{empty}

\null
 \vfill
\begin{center} 
{\LARGE Quantized gravito-electro-magnetic interactions of bilinear type\par}%
   \vskip 3em
   {\large C. Pierre
\\[5mm]
Institut de Mathématique pure et appliquée\\
Université de Louvain\\
Chemin du Cyclotron, 2\\
B-1348 Louvain-la-Neuve\\ Belgium\\
pierre@math.ucl.ac.be\par}

\end{center}

\vfill\eject

\null\vfill\vfill

{\begin{abstract}{\noindent 
\\
The interactions inside the (bisemi)particles and between them are of bilinear type and are envisaged from two points of view:
\vskip 11pt

The first approach, based on the reducible representations of algebraic bilinear semigroups, allows to describe explicitly the interactions between (bisemi)particles by means of gravito-electro-magnetic fields of interaction while the second approach, based on the consideration of (bi)connections, leads to Maxwell extended equations involving the gravitational field.
}

\end{abstract}

\vfill\eject} 
}

\setcounter{page}{1}

\section{Introduction}

\addtocontents{toc}{\protect\thispagestyle{empty}}

As the {\bbf elementary particles are considered\/} in the algebraic quantum theory (AQT) \cite{Pie2}, \cite{Pie4} {\bbf as being elementary bisemiparticles\/}:
\Bi
\item {\bf having  a twofold nature\/} based on the product of a left semiparticle localized in the upper half space by its symmetric right equivalent localized in the lower half space

\hspace{-5mm} and

\item {\bbf characterized by a three level embedded internal space-time structure\/},
\Ei
it is natural to envisage that {\bbf the interactions\/} inside the elementary (bisemi)particles and between them {\bbf are of bilinear type\/}.
\vskip 11pt

This contrasts obviously with {\bbf the linear treatment of the interactions\/} between elementary particles considered {\bbf in quantum field theories\/} and, especially, in gauge theories, which is {\bbf at the origin of the divergences\/} encountered in these theories as developed in this paper.

Einstein \cite{Ein3} pointed out previously that ``linear laws have solutions which satisfy the superposition principle but they do not describe the interactions between elementary particles''.
\vskip 11pt

It will be shown that the bilinear interactions between left and right internal structures of $J$ elementary bisemiparticles can be judiciously described in the bilinear frame of the global program of Langlands by the reducible representation of the algebraic general bilinear semigroup $\GL_{2J}(L_{\o v}\times L_v)$ of order $2J$ which provides the structure of the searched interaction field.
\vskip 11pt

At this stage, it is important to notice:
\Bi
\item that {\bbf the interactions in QFT are handled linearly by means of gauge theories\/} (dealing with the groups of transformations on the fields leaving invariant the Lagrangian density) {\bbf in the mathematical frame of the Erlanger program\/} of which studies have concerned the actions of the symmetry groups in geometry

\item while {\bbf the interactions in AQT are envisaged bilinearly throughout the functional representations of the algebraic bilinear semigroups in the framework of the Langlands global program\/} \cite{Pie3}.
\Ei
\vskip 11pt

In this context, the internal structure of elementary bisemiparticles is recalled in {\bf chapter 2\/} and it is proved that the interactions inside these are generated by (gravito)-electro-magnetic fields.

First of all, {\bbf the most internal space-time structure of the vacuum of an elementary (bisemi)fermion\/} $e^-$, $u^+$ and $d^-$ of the first family is developed: it {\bbf is an operator valued string field  $(\widetilde M^{T_p-S_p}_{ST_R}
\otimes \widetilde M^{T_p-S_p}_{ST_L})$ at this ``$ST$'' level\/} decomposing into:
\begin{multline*}
(\widetilde M^{T_p-S_p}_{ST_R}\otimes \widetilde M^{T_p-S_p}_{ST_L})\\
= [(\widetilde M^{T_p}_{ST_R}\otimes_D \widetilde M^{T_p}_{ST_L})
\oplus (\widetilde M^{S_p}_{ST_R}\otimes_D \widetilde M^{S_p}_{ST_L})]
\oplus (\widetilde M^{S_p}_{ST_R}\otimes_m \widetilde M^{S_p}_{ST_L})
\oplus (\widetilde M^{(T_p)-S_p}_{ST_R}\otimes_e \widetilde M^{(S_p)-T_p}_{ST_L})
\end{multline*}
where:
\Bi
\item $(\widetilde M^{T_p}_{ST_R}\otimes_D \widetilde M^{T_p}_{ST_L})$ is the ``time'' string field
\Bi
\item[-] composed of bistrings (i.e. (diagonal) products of right strings by their left correspondents) characterized by increasing ranks,
\item[and -] being given by a (bisemi)sheaf of differentiable bifunctions over the representation space of the bilinear algebraic semigroup $\GL_2(L_{\o v}\times L_v)_t$ (we refer to chapter 2 for the precise mathematical terminology);
\Ei

\item $(\widetilde M^{S_p}_{ST_R}\otimes_D \widetilde M^{S_p}_{ST_L})$ is the ``space'' string field of the ``$ST$'' level localized in  a space orthogonal to the ``time'' string field;

\item $(\widetilde M^{S_p}_{ST_R}\otimes_m \widetilde M^{S_p}_{ST_L})$ is the {\bbf magnetic string field responsible for the magnetic moment\/} of the bisemifermion at this ``$ST$'' level and generated by the exchange of magnetic biquanta between the right and left semifields $\widetilde M^{S_p}_{ST_R}$ and $\widetilde M^{S_p}_{ST_L}$;

\item $(\widetilde M^{(T_p)-S_p}_{ST_R}\otimes_e \widetilde M^{(S_p)-T_p}_{ST_L})$ is {\bbf the electric string field\/} at this ``$ST$'' level {\bbf responsible for the electric charge a this level\/} and generated by the exchange of electric biquanta.
\Ei
\vskip 11pt

Afterwards, the (operator) valued string field at the ``$ST$'' level, submitted to strong fluctuations, generates the {\bbf two enveloping middle ground ($MG$) and mass ($M$) string fields\/} $(\widetilde M^{T_p-S_p}_{MG_R}\otimes \widetilde M^{T_p-S_p}_{MG_L})$ and $(\widetilde M^{T_p-S_p}_{M_R}\otimes \widetilde M^{T_p-S_p}_{M_L})$ in such a way that {\bbf the interactions inside an elementary bisemifermion are crudely given by\/}:
\Bean
\item {\bbf the magnetic string fields at the ``$ST$'', ``$MG$'' and ``$M$'' levels\/} responsible for the corresponding magnetic moments;

\item {\bbf the electric string fields at these levels ``$ST$'', ``$MG$'' and ``$M$''\/} responsible for the electric charges on these shells.
\Ee
\vskip 11pt

Remark that the value of the electric charge at the ``$ST$'', ``$MG$'' or ``$M$'' level is the sum, over all sets of exchanged electric biquanta, of the pseudo-norms of the generators of the corresponding Lie algebras.
\vskip 11pt

It is also recalled that {\bbf the interactions inside a (bisemi)photon at $\gamma $ biquanta are provided by the exchanges of magnetic biquanta\/} generating magnetic subfields at the ``$ST$'', ``$MG$'' and ``$M$'' levels.

{\bbf Finally, the internal structure of bisemibaryons\/} is  reviewed: it {\bbf is characterized by the existence of a left and right ``core'' time semifield\/} (at the ``$ST$'', ``$MG$'' and ``$M$'' levels) {\bbf from which the ``time'' string semifields of the three left and right semiquarks are generated\/} in such a way that {\bbf a left and a right semibaryon\/} of a given bisemibaryon {\bbf interact\/} at the
``$ST$'', ``$MG$'' and ``$M$'' levels {\bbf by means of\/}:
\Bena
\item {\bbf the electric charges and the magnetic moments of the three bisemiquarks\/};

\item {\bbf a gravito-electro-magnetic field resulting from the bilinear interactions between the left and right semiquarks of different bisemiquarks\/};

\item {\bbf a strong gravitational and electric field resulting from the bilinear interactions between the central core structures of the left and right semibaryons and, respectively, the right and left semiquarks\/}.
\Ee
\vskip 11pt

{\bbf Chapter 3\/} is devoted to the study of the bilinear interactions between bisemiparticles according to two points of view.  The first approach, based on the reducible representation of the algebraic bilinear semigroup $\GL_{2J}(L_{\o v}\times L_v)$, allows to describe explicitly the interactions between ``$J$'' bisemiparticles by means of gravito-electro-magnetic fields of interaction while the second approach, based on the consideration of (bi)connections, leads to Maxwell extended equations in a bilinear mathematical frame.

More precisely, {\bbf the\/} (operator valued) {\bbf string fields\/} at the
``$ST$'', ``$MG$'' or ``$M$'' level, {\bbf of a set of $J$ interacting bisemiparticles are given in the first approach by\/}:
\[ (\widetilde M^{T_p-S_p}_R {(J)}\otimes \widetilde M^{T_p-S_p}_L {(J)})
= \bigoplus^J_{i=1} (\widetilde M^{T_p-S_{p}}_R {(i)}\otimes \widetilde M^{T_p-S_{p}}_L {(i)})
\ \bigoplus^J_{i\neq j=1} (\widetilde M^{T_p-S_{p}}_R {(i)}\otimes \widetilde M^{T_p-S_{p_j}}_L {(j)})\]
{\bf where the mixed direct sum $ \bigoplus^J_{i\neq j=1} (\widetilde M^{T_p-S_{p}} _R{(i)}\otimes \widetilde M^{T_p-S_{p}}_L {(j)})$ refers to the bilinear interaction fields\/} between the right and left semiparticles belonging to different bisemiparticles.
\vskip 11pt

{\bbf These interaction fields are\/}:
\Bean
\item in the case of (bisemi)leptons: of gravitational, electric and magnetic nature (section 3.4),
\item in the case of (bisemi)photons: of gravitational and magnetic nature (section 3.5),
\item in the case of (bisemi)baryons: of
\Be 
\item strong gravitational nature between right and left different central cores,
\item gravitational, electric and magnetic nature between right and left different semiquarks,
\item strong gravitational and electric  nature between \rl core time structures and \lr semiquarks (sections 3.6 to 3.8).
\Ee
\Ee
\vskip 11pt

In the second approach at the ``$M$'' level, {\bbf the mass bioperator\/} $(\MM_R\otimes \MM_L)$, acting on every mass bisection or bistring of an elementary bisemifermion and {\bbf endowed with the infinitesimal biconnection $((e)A_R\otimes (e)A_L)$, develops according to\/}:
\[ ((\MM_R+A_R) \otimes (\MM_L+A_L))
= (\MM_R\otimes \MM_L)+(A_R\otimes A_L) 
+ [(\MM_R\otimes A_L)+(A_R\otimes \MM_L)]
\]
{\bbf where $[(\MM_R\otimes A_L)+(A_R\otimes \MM_L)]$ is the gravito-electro-magnetic interaction field operator of which tensorial form is the $\MM A_{mn}$ symmetric gravito-electro-magnetic tensor\/} corresponding to the $\FF_{mn}$ antisymmetric tensor of electro-magnetism.

So, the introduction of bilinearity in AQT gives rise to the gravito-electro-magnetism leading to {\bbf a powerful unification of  gravitation with  electromagnetism\/}, as hoped by A. Einstein.
\vskip 11pt

In this new context, the condition of $4D$-null divergence $\partial^n\MM A_{mn}=0$ applied to the GEM tensor $\MM A_{mn}$ leads to a pair of GEM (gravito-electro-magnetic) differential equations which describe not more, as in the Maxwell equations, the flux of $\vec E$ through a closed surface $(\vec\nabla\centerdot\vec E)$ or the circulation of $\vec B$ around a  loop, respectively by means of the charge density inside and current through the loop, but, in function of the variation in time of the time gravitational field $G_t$ and of the flux of the space gravitational field $G$ through the loop.
\vskip 11pt

{\bbf In chapter 4, the Feynman paths\/} relative to all kinds of gravito-electro-magnetic interactions in AQT {\bbf are reinterpreted on the basis of the following considerations\/}:
\Bean
\item {\bbf The exchange of gravitational biquanta\/}, responsible for the gravitational force, is considered on an equal footing as the exchange of electro-magnetic biquanta corresponding to the virtual photons of QED.

\item {\bbf The interactions\/} between fields {\bbf are not realized\/} in AQT at a fundamental level {\bf by perturbative series\/}.

\item   The transition amplitudes of the Feynmann paths in QFT become {\bbf transition intensities in AQT\/} due to the bilinear character of this theory.
\Ee
\vskip 11pt

As each ``$ST$'', ``$MG$'' or ``$M$'' (operator-valued) field of space or time is given in AQT by the (sum of the) set of products, right by left, of its sections which are strings, {\bbf the only basic diagrams of interaction\/} are those involving exchanges of gravitational, electric and magnetic biquanta between left and right strings: they {\bbf correspond to the first order diagrams of Feynmann\/}.

In the philosophy of AQT, the left and right strings are one-dimensional waves, homotopic to circles $S^1$, having two senses of rotation, in such a way that:
\Bean
\item {\bbf the Feynmann path intervals are reinterpreted as arcs of circles\/} measured by angles of rotation,

\item the \lr {\bbf Green's propagator\/} or a \lr string {\bbf is given by the one parameter group of diffeomorphisms\/} shifting each point of the string by a small interval (of arc).

\item {\bbf the different paths\/} from a point $A$ to a point $B$ {\bbf correspond to the possible normal modes of the exchanged biquanta\/}.
\Ee
\vskip 11pt

In consequence, {\bbf AQT is\/} a mathematical theory {\bbf exempt of divergences\/}: for example, the famous self-energy diagram in QFT is then replaced by a diagram involving the exchange of one magnetic biquantum inside a bistring.

\section[Internal gravito-electromagnetic interactions of (elementary) particles]{Internal gravito-electromagnetic interactions of\linebreak (elementary) particles}

The aim of this chapter consists in:
\Bean
\item recalling what is the {\bbf internal structure of the elementary particles\/} which are  described in the present context as bisemiparticles, a bisemiparticle being given by the product of a left semiparticle localized in the upper half space by its symmetric right equivalent localized in the lower half space.

\item showing that an elementary bisemiparticle sticks together by means of the {\bbf bilinear (gravito)-electro-magnetic fields of interaction\/} between its right and left components.
\Ee
\vskip 11pt

\subsection{The classification of particles in AQT}

The classification of elementary (bisemi)particles adopted in this algebraic quantum theory corresponds to the standard classification used in quantum field theories, excepts perhaps with regard to some gauge bosons of the non abelian gauge field theories (since AQT is not essentially a gauge theory but ``covers'' in some way the gauge fields theories).
\vskip 11pt

In this respect, the {\bbf elementary (bisemi)fermions\/} are:
\Bi
\item the leptons $e^-$, $\mu ^-$, $\tau ^-$ and their neutrinos;
\item the quarks $u^+$, $d^-$, $s^-$, $c^+$, $b^-$, $t^+$;
\Ei
and the {\bbf elementary (bisemi)bosons\/} are:
\Bi
\item the photons;
\item the gravitational, electronic and magnetic (bisemi)bosons.
\Ei
\vskip 11pt

The {\bbf hadrons\/} have to be added to this list, although they are not essentially elementary since they are composed from a set of three quarks ion the case of baryons and of two quarks in the case of mesons.
\vskip 11pt

First of all, {\bbf the space-time structure of\/} the elementary bisemifermions {\bbf $e^-$, $u^+$ and $d^-$\/} of the first family {\bf will be recalled\/}.

In order to take into account the main features of quantum field theories \cite{Pie4}, it was assumed that the {\bbf mass shell\/} of an elementary bisemifermion {\bbf could be generated from\/} its most internal space-time structure interpreted as {\bbf its internal vacuum structure\/}.
\vskip 11pt

\subsection{\bbf The space-time vacuum structure of $e^-$, $u^+$ and $d^-$}

\Bi
\item The internal space structure $(M^S_{ST_R}\otimes M^S_{ST_L})$ of the vacuum of an elementary bisemifermion is generated from the corresponding {\bbf vacuum time structure $(M^T_{ST_R}\otimes M^T_{ST_L})$\/} which is given \cite{Pie5} by the representation space $\Repsp(\GL_2(L_{\o v}\times L_v)_t)$ of the {\bbf bilinear algebraic semigroup $\GL_2(L_{\o v}\times L_v)_t$\/} over the product $(L_{\o v}\times L_v)$ of the sets of completions \cite{B-T}, \cite{Che}, \cite{Har}, \cite{Lan}, \cite{Ser}
\begin{align*}
L_{v} &= \{L_{v_1},\dots,L_{v_{\mu ,m_\mu }},\dots,L_{v_{q,m_q}}\}\\
\text{and} \quad
L_{\o v} &= \{L_{\o v_1},\dots,L_{\o v_{\mu ,m_\mu }},\dots,L_{\o v_{q,m_q}}\}\;.\end{align*}
\vskip 11pt

\item The set of {\bbf equivalent completions\/} $\{L_{v_{\mu ,m_\mu }}\}_{m_\mu }$
\resp{$\{L_{\o v_{\mu ,m_\mu }}\}_{m_\mu }$} of the $\mu $-th real place $v_\mu $ \resp{$\o v_\mu $}, $1\le \mu \le q$, is in one-to-one correspondence with the set of real pseudo-ramified algebraic extensions $\{F_{v_{\mu ,m_\mu }}\}_{m_\mu }$
\resp{$\{F_{\o v_{\mu ,m_\mu }}\}_{m_\mu }$} of a global number field $K$ of characteristic 0 and is characterized by a rank  
\[[L_{v_{\mu ,m_\mu }}:K]\equiv [L_{\o v_{\mu ,m_\mu }}:K]=*+\mu \centerdot N\]
equal to the {\bbf Galois extension degree\/}  of $F_{v_{\mu ,m_\mu }}$ \resp{$F_{\o v_{\mu ,m_\mu }}$}
 
 in such a way that $[L_{v_{\mu ,m_\mu }}:K]$ is an integer modulo $N$ where $*$ denotes an integer inferior to $N$.
 
 {\bbf The integer $N$\/} is the Galois extension degree of the irreducible algebraic closed subsets interpreted as time quanta.
 \vskip 11pt
 
 \item The $(\mu ,m_\mu )$-th conjugacy class representative $\GL_2(L_{\o v_{\mu ,m_\mu }}\times L_{v_{\mu ,m_\mu }})_t$ of $\GL_2(L_{\o v}\times L_v)_t$ has for representation the
 $\GL_2(L_{\o v_{\mu ,m_\mu }}\times L_{v_{\mu ,m_\mu }})$-subbisemimodule
 $(M_{L_{\o v_{\mu ,m_\mu }}}\otimes M_{L_{v_{\mu ,m_\mu }}})$ which is such that
 $M_{L_{\o v_{\mu ,m_\mu }}}$ rotates in opposite sense with respect to its symmetric component $M_{L_{v_{\mu ,m_\mu }}}$ in the corresponding Lie algebra representative $\gl_2
 (L_{\o v_{\mu ,m_\mu }}\times L_{v_{\mu ,m_\mu }})$.
 
 {\bbf The set of conjugacy class representatives of $\GL_2(L_{\o v}\times L_v)_t$\/} can be set up into a tower \cite{G-D}, \cite{Bour}, \cite{Weil}:
\[
 \GL_2(L_{\o v_1}\times L_{v_1}) \subset \dots \subset
 \GL_2(L_{\o v_{\mu ,m_\mu }}\times L_{v_{\mu ,m_\mu }}) 
\subset \dots \subset
 \GL_2(L_{\o v_{q ,m_q }}\times L_{v_{q ,m_q }}) \]
 of which components are characterized by increasing ranks.
 \vskip 11pt
 
 \item It was seen in \cite{Pie2} and in \cite{Pie5} that,

{\bbf under the composition of maps $\gamma _{t\RL\to r\RL}\circ E\RL$}, where
 \Bean
 \item {\bbf $E\RL$}  : \quad $\Repsp (\GL_2(L_{\o v}\times L_v))_t $
 
 \hfill $\To\Repsp (\GL^*_2(L^*_{\o v}\times L^*_v)_t )
 \oplus  \Repsp (\GL^I_2(L^I_{\o v}\times L^I_v)_t )$
 
 {\bbf is a smooth biendomorphism\/} transforming the representation space\linebreak
 $\Repsp (\GL_2(L_{\o v}\times L_v)_t )\equiv (M^T_{ST_R}\otimes M^T_{ST_L})$
 of $\GL_2(L_{\o v}\times L_v)_t$ into a reduced ``time'' representation space
 $\Repsp (\GL^*_2(L^*_{\o v}\times L^*_v)_t)$ over a set of products of reduced 
completions $(L^*_{\o v}\times L_v)_t$ and into a
complementary disconnected ``time'' representation space
 $\Repsp (\GL^I_2(L^I_{\o v}\times L^I_v)_t)$ in such a way that:
 \begin{align*}
 [L_{v_{\mu ,m_\mu }}:K] 
&= [L^*_{v_{\mu ,m_\mu }}:K] + [L^I_{v_{\mu ,m_\mu }}:K]\\
 \rresp{[L_{\o v_{\mu ,m_\mu }}:K] &= [L^*_{\o v_{\mu ,m_\mu }}:K] + [L^I_{\o v_{\mu ,m_\mu }}:K]}\\
 & \hspace{-1.5cm}
 \forall\ L_{v_{\mu ,m_\mu }}\in L_v\ , \; 
L^*_{v_{\mu ,m_\mu }}\in L^*_v\ , \; 
 L^I_{v_{\mu ,m_\mu }}\in L^I_v\ , \; 
 L^I_{\o v_{\mu ,m_\mu }}\in L^I_{\o v}\;.
\end{align*}
 \vskip 11pt
 
 \item {\bbf $\gamma _{t\RL\to r\RL}$}: \quad 
$\Repsp (\GL^I_2(L^I_{\o v}\times L^I_v)_t )\To
\Repsp (\GL^I_2(L^I_{\o v}\times L^I_v)_r )$

sends the complementary disconnected ``time'' representation space\linebreak
$\Repsp (\GL^I_2(L^I_{\o v}\times L^I_v)_t )$ into the orthogonal representation space\linebreak
$\Repsp (\GL^I_2(L^I_{\o v}\times L^I_v)_r )$ which is of ``spatial'' nature,
\Ee

{\bbf the ``time'' representation space\/}
$\Repsp (\GL^I_2(L^I_{\o v}\times L^I_v)_t )$ {\bbf could be transformed into\/}:
\Bena
\item {\bf a reduced ``time'' representation space\/}
$\Repsp (\GL^*_2(L^*_{\o v}\times L^*_v)_t )$
\Ee
{\bf and into}
\Be
\stepcounter{enumii}
\item {\bf 	 a complementary orthogonal ``space'' representation space\/}\linebreak
$\Repsp (\GL^I_2(L^I_{\o v}\times L^I_v)_r )$
\Ee
\Ei
\vskip 11pt

\subsection[The (operator valued) string field of the $ST$-level]{\bbf The (operator valued) string field of the $ST$-level}

\Bi
\item The set $\{\phi _L(M_{L_{v_{\mu ,m_\mu }}})\}_{\mu ,m_\mu }$
\resp{$\{\phi _R(M_{L_{\o v_{\mu ,m_\mu }}})\}_{\mu ,m_\mu }$}
of {\bbf $\cit$-valued differentiable functions\/} on 
$\{M_{L_{v_{\mu ,m_\mu }}}\}$ \resp{$\{M_{L_{\o v_{\mu ,m_\mu }}}\}$},
localized in the upper (resp. lower) half space and defined over the $T_2(L_v)_t$ \resp{$T^t_2(L_{\o v})_t$}-semimodule $M^T_{ST_L}$ \resp{$M^T_{ST_R}$}, constitutes the set $\Gamma (\widetilde M^T_{ST_L})$
\resp{$\Gamma (\widetilde M^T_{ST_R})$} of sections of the semisheaf of rings
$\widetilde M^T_{ST_L}$ \resp{$\widetilde M^T_{ST_R}$}.
\vskip 11pt

\item {\bbf The set of differentiable bifunctions\/}
$\{\phi _R(M_{L_{\o v_{\mu ,m_\mu }}}) \otimes \phi _L(M_{L_{v_{\mu ,m_\mu }}})\}$ over the
$\GL_2(L_{\o v}\times L_v)_t$-bisemimodule $(M^T_{ST_R}\otimes M^T_{ST_L})$ constitutes the set of bisections of the bisemisheaf of rings
$(\widetilde M^T_{ST_R}\otimes \widetilde M^T_{ST_L})$.
\vskip 11pt

 \item The bisemisheaf of rings
$(\widetilde M^T_{ST_R}\otimes \widetilde M^T_{ST_L})$ is a
{\bf physical ``time'' string field of the internal vacuum\/} of an elementary bisemifermion of the first family.
\vskip 11pt

\item Similarly, the bisemisheaf of rings
$(\widetilde M^S_{ST_R}\otimes \widetilde M^S_{ST_L})$ on the ``space'' complementary representation space
$\Repsp (\GL^I_2(L^I_{\o v}\times L^I_v)_r )$ is the {\bbf ``space'' string field of the internal vacuum\/} of an elementary bisemifermion of the first family.
\vskip 11pt

\item {\bf The ``time'' and ``space'' string fields of the internal vacuum\/} of an elementary bisemifermion is then given by:
\begin{multline*}
\widetilde M^{T-S}_{ST_R} \otimes \widetilde M^{T-S}_{ST_L}\\
\begin{aligned}
&= (\widetilde M^{T}_{ST_R} \oplus \widetilde M^{S}_{ST_R} )
\otimes (\widetilde M^{T}_{ST_L} \oplus \widetilde M^{S}_{ST_L}) \\
&= (\widetilde M^{T}_{ST_R} \otimes \widetilde M^{T}_{ST_L})
\oplus (\widetilde M^{S}_{ST_R} \otimes \widetilde M^{S}_{ST_L})
\oplus [ (\widetilde M^{T}_{ST_R} \otimes \widetilde M^{S}_{ST_L})
\oplus   (\widetilde M^{S}_{ST_R} \otimes \widetilde M^{T}_{ST_L})]
\end{aligned}\end{multline*}
where {\bbf the cross product string fields\/}, also noted
$(\widetilde M^{T}_{ST_R} \otimes_e \widetilde M^{S}_{ST_L})$
and\linebreak $(\widetilde M^{S}_{ST_R} \otimes_e \widetilde M^{T}_{ST_L})$, {\bbf are responsible for the electric charge\/} of this bisemifermion at the ``$ST$''-internal vacuum level, as it will be seen in the following.
\vskip 11pt

\item {\bbf The operator-valued string field $(\widetilde M^{T_p-S_p}_{ST_R} \otimes \widetilde M^{T_p-S_p}_{ST_L})$ of the $ST$-level\/} of the internal vacuum is a (perverse) bisemisheaf obtained from $(\widetilde M^{T-S}_{ST_R} \otimes \widetilde M^{T-S}_{ST_L}
)$ by the action of the differential bioperator
\begin{multline*}
T^{T-S}_{R;ST} \otimes T^{T-S}_{L;ST}\\
= \L(-i\ \F{\hbar_{ST}}{c_{t\to r;ST}}
\ \L\{ s_{0R}\ dt_0;s_{x_R}\ dx,s_{y_R}\ dy,s_{z_R}\ dz\R\}\R)\\
\otimes \L(+i\ \F{\hbar_{ST}}{c_{t\to r;ST}}
\ \L\{ s_{0L}\ dt_0;s_{x_L}\ dx,s_{y_L}\ dy,s_{z_RL}\ dz\R\}\R)
\end{multline*}
on every bisection of $(\widetilde M^{T-S}_{ST_R} \otimes \widetilde M^{T-S}_{ST_L})$ leading to:
\[ T^{T-S}_{R;ST} \otimes T^{T-S}_{L;ST}: \quad
(\widetilde M^{T-S}_{ST_R} \otimes  \widetilde M^{T-S}_{ST_L})\To
(\widetilde M^{T_p-S_p}_{ST_R} \otimes \widetilde M^{T_p-S_p}_{ST_L}) \]
as it was developed in \cite{Pie5}.
\vskip 11pt

\item {\bbf Each \lr section\/} of a bisection of the operator-valued string field
$(\widetilde M^{T_p}_{ST_R} \otimes \widetilde M^{T_p}_{ST_L})$ of ``time'' {\bbf has two possible senses of rotation\/} due to the directional differential $\vec s_{0L}\ dt_0$
\resp{$\vec s_{0R}\ dt_0$}.  

Similarly, each \lr section of a bisection of the operator-valued string field
$(\widetilde M^{S_p}_{ST_R} \otimes \widetilde M^{S_p}_{ST_L})$ of ``space''  has two possible senses of rotation  due to the directional differential $\vec s_{r_L}\ d\vec r$
\resp{$\vec s_{r_R}\ d\vec r$} with $d\vec r=\{dx,dy,dz\}$.
\vskip 11pt

\item Remark that:
\Bean
\item  the senses of rotation to the \lr sections of
$\widetilde M^{S_p}_{ST_L}$ \resp{$\widetilde M^{S_p}_{ST_R}$} are directly related to the
{\bf spin of the considered (bisemi)fermion\/}.
\item {\bbf Each bisection\/} of $(\widetilde M^{S_p}_{ST_R}\otimes \widetilde M^{S_p}_{ST_L})$ {\bbf is a bistring\/} given by the product of a right section of
$\widetilde M^{S_p}_{ST_R}$ by the corresponding symmetric left section of
$\widetilde M^{S_p}_{ST_L}$.
\item {\bbf Each internal field\/}, for example
$(\widetilde M^{S_p}_{ST_R}\otimes \widetilde M^{S_p}_{ST_L})$, {\bbf is composed of\/}
a set of bisections which are {\bbf bistrings behaving like harmonic oscillators\/}.
\Ee
\Ei

\subsection[(Operator valued) string field of the ``$ST$'' level of an elementary bisemifermion]{\bbf (Operator valued) string field of the ``$ST$'' level of an elementary bisemifermion}

It was seen in section 2.3 that the (operator valued) string field of the internal vacuum of an elementary bisemifermion at the ``$ST$'' level is given by:
\begin{multline*}
 \widetilde M^{T_p-S_p}_{ST_R}\otimes \widetilde M^{T_p-S_p}_{ST_L}\\
= (\widetilde M^{T_p}_{ST_R}\otimes \widetilde M^{T_p}_{ST_L})
\oplus (\widetilde M^{S_p}_{ST_R}\otimes \widetilde M^{S_p}_{ST_L})
+ [(\widetilde M^{T_p}_{ST_R}\otimes_e \widetilde M^{S_p}_{ST_L})
+ (\widetilde M^{S_p}_{ST_R}\otimes_e \widetilde M^{T_p}_{ST_L}]\end{multline*}
in such a way that:
\Bean
\item $(\widetilde M^{T_p}_{ST_R}\otimes \widetilde M^{T_p}_{ST_L})$ is {\bbf the ``time'' string field of the ``$ST$'' level\/} composed of packets of bistrings behaving like harmonic oscillators.  Consequently, the tensor product ``$\otimes$'' is a diagonal tensor product, written ``$\otimes_D$'', characterized by a diagonal bilinear basis \cite{Pie2}.

So, $(\widetilde M^{T_p}_{ST_R}\otimes \widetilde M^{T_p}_{ST_L})$ will be rewritten according to $(\widetilde M^{T_p}_{ST_R}\otimes_D \widetilde M^{T_p}_{ST_L})$.

\item $(\widetilde M^{S_p}_{ST_R}\otimes \widetilde M^{S_p}_{ST_L})$ is {\bbf the ``space'' string field of the ``$ST$'' level\/} composed of packets of bistrings which can be compactified on a three-dimensional (bilinear) semimanifold as described in \cite{Pie5}.

On the other hand, as these bistrings are rotating, they are submitted to a Coriolis (bi)force responsible for {\bbf magnetic smooth biendormophisms} $E_{R_{\mu ,m_\mu }}\otimes_m E_{L_{\mu ,m_\mu }}$:
\begin{multline*}
 E_{R_{\mu ,m_\mu }}\otimes_m E_{L_{\mu ,m_\mu }}:
\quad \phi ^{S_p}_R(M_{L_{\o v_{\mu ,m_\mu }}}) \otimes \phi ^{S_p}_L(M_{L_{v_{\mu ,m_\mu }}}) \\
\To (\phi ^{S_p}_R(M_{L_{\o v_{\nu ,m_\nu }}}) \otimes_D
\phi ^{S_p}_L(M_{L_{v_{\nu ,m_\nu }}}))
\bigoplus^\rho _{k=1} (\widetilde M^I_{k_{R;ST}} \otimes _m \widetilde M^I_{k_{L;ST}})\;,
\end{multline*}
as developed in \cite{Pie2},

{\bbf in such  a way that $\rho =\mu -\nu $\/}, $\rho \in\NN$, {\bbf magnetic biquanta\/} are taken away from the bistring
$\phi ^{S_p}_R(\widetilde M_{L_{\o v_{\mu ,m_\mu }}}) \otimes
\phi ^{S_p}_L(M_{L_{v_{\mu ,m_\mu }}})$, or, more exactly, {\bbf are discretely exchanged\/} between the right and left string
$\phi ^{S_p}_R(M_{L_{\o v_{\mu ,m_\mu }}})$ and
$\phi ^{S_p}_L(M_{L_{v_{\mu ,m_\mu }}})$ as it will be described in chapter 4 by means of the Feynmann (bi)graphs.

Consequently, the bistring $(\phi ^{S_p}_R(M_{L_{\o v_{\mu ,m_\mu }}})\otimes
\phi ^{S_p}_L(M_{L_{v_{\mu ,m_\mu }}}))$ is turned into one with diagonal metric ``$\otimes_D$'' and ``$\nu $'' ``permanent'' diagonal biquanta.  These exchanged magnetic biquanta are thus responsible for the {\bbf magnetic moment\/} of the elementary bisemifermion at the ``$ST$'' level of its internal vacuum.

\item the cross product string field
$(\widetilde M ^{T_p}_{ST_R}\otimes_e \widetilde M ^{S_p}_{ST_L})
+(\widetilde M ^{S_p}_{ST_R}\otimes_e \widetilde M ^{T_p}_{ST_L})$, responsible for the cohesion of the ``$ST$'' level by {\bbf the exchange of electric biquanta\/}, is such that {\bbf only one of these string fields is activated\/} or really generated because:
\Bi
\item mathematically, as we have two symmetric cross products, {\bbf there is an obstruction\/} to the existence of the two ``simultaneous'' string fields.
\item physically, this would correspond to a symmetry breaking mechanism.
\Ei

So, the string field $(\widetilde M ^{T_p}_{ST_R}\otimes_e \widetilde M ^{S_p}_{ST_L})$ would be the string field responsible for a negative electric charge of the bisemifermion at the ``$ST$'' level while $(\widetilde M ^{S_p}_{ST_R}\otimes_e \widetilde M ^{T_p}_{ST_L})$ would be the string field responsible for a positive electric charge at the ``$ST$'' level, as developed in \cite{Pie2}.
\Ee
 
\subsection{Proposition}

{\em {\bfseries\boldmath The operator valued string field of the internal vacuum of an elementary bisemifermion at the $ST$ level splits according to:}
\begin{multline*}
 (\widetilde M^{T_p-S_p}_{ST_R}\otimes \widetilde M^{T_p-S_p}_{ST_L})\\
= (\widetilde M^{T_p}_{ST_R}\otimes_D \widetilde M^{T_p}_{ST_L})
\oplus (\widetilde M^{S_p}_{ST_R}\otimes_D \widetilde M^{S_p}_{ST_L})
\oplus (\widetilde M^{S_p}_{ST_R}\otimes_m \widetilde M^{S_p}_{ST_L})
\oplus (\widetilde M^{(T_p)-S_p}_{ST_R}\otimes_e \widetilde M^{(S_p)-T_p}_{ST_L})\end{multline*}
where:
\Bi
\item {\bfseries\boldmath $(\widetilde M^{T_p}_{ST_R}\otimes_D \widetilde M^{T_p}_{ST_L})$ is the ``time'' string field\/} of the ``$ST$'' level composed of bistrings characterized by increasing ranks.

\item {\bfseries\boldmath $(\widetilde M^{S_p}_{ST_R}\otimes_D \widetilde M^{S_p}_{ST_L})$ is the ``space'' string field\/} of the ``$ST$'' level, localized in a space orthogonal to
$(\widetilde M^{T_p}_{ST_R}\otimes \widetilde M^{T_p}_{ST_L})$ and composed of bistrings characterized by increasing ranks.

\item {\bfseries\boldmath $(\widetilde M^{S_p}_{ST_R}\otimes_m \widetilde M^{S_p}_{ST_L})$ is the ``magnetic'' string field\/} at the ``$ST$'' level, responsible of the magnetic moment of the bisemifermion at this level by the exchange of magnetic biquanta between the left and right semifields $\widetilde M^{S_p}_{ST_R}$ and  $\widetilde M^{S_p}_{ST_L}$ of ``space''.

\item {\bfseries\boldmath $(\widetilde M^{(T_p)-S_p}_{ST_R}\otimes_e \widetilde M^{(S_p)-T_p}_{ST_L})$ is the electric string field\/} at the ``$ST$'' level, responsible for the electric charge of the bisemifermion at this level by the exchange of electric biquanta between the left semifield of time (resp. space) and the right semifield of space (resp. time).
\Ei}

\bpr This proposition is a direct consequence of section 2.4. \epr

\subsection{Proposition}

{\em The internal vacuum ``$ST$'' structure of an elementary (bisemi)fermion $e^-$, $u^+$ or $d^-$ is characterized by:
\Bean
\item the numbers of bisections, i.e. bistrings, of its ``time'' and ``space'' string fields;
\item the sets of ranks of these bisections associated with the numbers of biquanta on these.
\Ee}
\vskip 11pt

\bpr
\Bean
\item The ``time'' string field $(\widetilde M^{T_p}_{ST_R} \otimes_D \widetilde M^{T_p}_{ST_L})$ is composed of a set of $q$ packets of\linebreak $m_q=\sup(m_q)$ bistrings, or bisections, in such  a way that the $(\mu ,m_\mu )$-th-bistring $\phi ^{T_p}_R(M_{L_{\o v_{\mu ,m_\mu }}} )\otimes\phi ^{T_p}_L(M_{L_{v_{\mu ,m_\mu }}} )$ be composed of $\mu $ biquanta.

Indeed, its rank $n_{\mu _{R-L}}$ is given by:
\begin{align*}
n_{\mu _{R-L}}
&= [L_{\o v_{\mu ,m_\mu }}:K] + [L_{v_{\mu ,m_\mu }}:K] \\
&= 2\ (*+\mu \centerdot N)\\
&\simeq 2\ \mu \centerdot N && \text{(if the zero class if only considered)}\end{align*}
and corresponds to $\mu $ biquanta at $N$ (bi)automorphisms of Galois according to section 2.2.

\item The ``space'' string field $(\widetilde M^{S_p}_{ST_R} \otimes_D \widetilde M^{S_p}_{ST_L})$ is generated from the ``time'' string field
$(\widetilde M^{T_p}_{ST_R} \otimes_D \widetilde M^{T_p}_{ST_L})$ by means of the
$(\gamma _{t\RL\to r\RL}\circ \EE\RL)$ morphism introduced in section 2.2 in such a way that:

if each bistring of time generates by emergence a bistring of space according to:
\begin{multline*}
\gamma _{t\RL\to r\RL}\circ \EE\RL: \quad \phi ^{T_p}_R(M_{L_{\o v_{\mu ,m_\mu }}})
\otimes \phi ^{T_p}_L(M_{L_{v_{\mu ,m_\mu }}})\\
\To (\phi ^{*T_p}_R(M_{L_{\o v_{\beta  ,m_\beta  }}}) \otimes
\phi ^{*T_p}_L(M_{L_{v_{\beta  ,m_\beta  }}}))\\
\oplus (\phi ^{S_p}_R(M_{L_{\o v_{\gamma   ,m_\gamma   }}}) \otimes
\phi ^{S_p}_L(M_{L_{v_{\gamma  ,m_\gamma  }}}))\;, \end{multline*}
then we have the following relation between their ranks:
\[ n_{\mu \RL}=n_{\beta \RL}+n_{\gamma \RL}\]
showing that the ``time'' bistring at ``$\mu $'' biquanta has been transformed into a reduced ``time'' bistring at ``$\beta $'' biquanta and into a complementary ``space'' bistring at ``$\gamma $'' biquanta.\epr
\Ee
\vskip 11pt

\subsection{The middle ground and mass string fields}

\Bi
\item It was seen in \cite{Pie2} and in \cite{Pie4} that the time and space string fields of the internal vacuum ``$ST$'' structure of an elementary bisemifermion were submitted to strong fluctuations generating on their sections degenerate singularities which are able to produce by versal deformations and blowups of these two new covering string fields: the middle ground and mass string fields embedding the internal vacuum ``$ST$'' string field according to:
\[ (\widetilde M^{T_p-S_p}_{ST_R} \otimes \widetilde M^{T_p-S_p}_{ST_L})
\subset (\widetilde M^{T_p-S_p}_{MG_R} \otimes \widetilde M^{T_p-S_p}_{MG_L})
\subset (\widetilde M^{T_p-S_p}_{M_R} \otimes \widetilde M^{T_p-S_p}_{M_L})\;.\]

\item The middle ground and mass string fields
$(\widetilde M^{T_p-S_p}_{MG_R} \otimes \widetilde M^{T_p-S_p}_{MG_L})$
and $(\widetilde M^{T_p-S_p}_{M_R} \otimes \widetilde M^{T_p-S_p}_{M_L})$
give also rise to a magnetic string field and to an electric string field by the same splitting considered for the ``$ST$'' string field in proposition 2.5.
\Ei
\vskip 11pt

\subsection{Proposition}
{\em If the inteactions between the ``$ST$'', ``$MG$'' and ``$M$'' string fields are assumed to be negligible, then {\bfseries the internal structure of an elementary bisemifermion will be given by the superposition of the following string fields\/}:
\begin{multline*}
(\widetilde M^{T_p-S_p}_{ST_R} 
\oplus \widetilde M^{T_p-S_p}_{MG_R}
\oplus \widetilde M^{T_p-S_p}_{M_R})
\otimes 
(\widetilde M^{T_p-S_p}_{ST_L} 
\oplus \widetilde M^{T_p-S_p}_{MG_L}
\oplus \widetilde M^{T_p-S_p}_{M_L})\\
= [
(\widetilde M^{T_p-S_p}_{ST_R} \otimes_D \widetilde M^{T_p-S_p}_{ST_L})
\oplus (\widetilde M^{T_p-S_p}_{MG_R} \otimes_D \widetilde M^{T_p-S_p}_{MG_L})
\oplus (\widetilde M^{T_p-S_p}_{M_R} \otimes_D \widetilde M^{T_p-S_p}_{M_L})]\qquad \\
\begin{aligned}
&\oplus [ 
(\widetilde M^{S_p}_{ST_R} \otimes_m \widetilde M^{S_p}_{ST_L})
\oplus (\widetilde M^{S_p}_{MG_R} \otimes_m \widetilde M^{S_p}_{MG_L})
\oplus (\widetilde M^{S_p}_{M_R} \otimes_m \widetilde M^{S_p}_{M_L})
]
\\
&\oplus [ 
(\widetilde M^{(T_p)-S_p}_{ST_R} \otimes_e \widetilde M^{(S_p)-T_p}_{ST_L})
\oplus (\widetilde M^{(T_p)-S_p}_{MG_R} \otimes_e \widetilde M^{(S_p)-T_p}_{MG_L})
\oplus (\widetilde M^{(T_p)-S_p}_{M_R} \otimes_e \widetilde M^{(S_p)-T_p}_{M_L})
]\end{aligned}
\end{multline*}
where:
\Bean
\item $(\widetilde M^{T_p-S_p}_{ST_R} \otimes_D \widetilde M^{T_p-S_p}_{ST_L})$,
$(\widetilde M^{T_p-S_p}_{MG_R} \otimes_D \widetilde M^{T_p-S_p}_{MG_L})$
and $(\widetilde M^{T_p-S_p}_{M_R} \otimes_D \widetilde M^{T_p-S_p}_{M_L})$ are {\bfseries\boldmath the ``diagonal'' ``$ST$'', ``$MG$'' and ``$M$'' string fields\/} composed of packets of bistrings with increasing numbers of biquanta in such a way that the $\mu $-th packet of closed bistrings at the ``$ST$'' level is successively covered by the corresponding $\mu $-packet of open bistrings at the middle ground (``$MG$'') and mass (``$M$'') levels.

\item $(\widetilde M^{S_p}_{ST_R} \otimes_m \widetilde M^{S_p}_{ST_L})$,
$ (\widetilde M^{S_p}_{MG_R} \otimes_m \widetilde M^{S_p}_{MG_L})$
and $ (\widetilde M^{S_p}_{M_R} \otimes_m \widetilde M^{S_p}_{M_L})$ are {\bfseries\boldmath the magnetic string fields\/} at the ``$ST$'', ``$MG$'' and ``$M$'' levels.

\item $(\widetilde M^{(T_p)-S_p}_{ST_R} \otimes_e \widetilde M^{(S_p)-T_p}_{ST_L})$,
$(\widetilde M^{(T_p)-S_p}_{MG_R} \otimes_e \widetilde M^{(S_p)-T_p}_{MG_L})$
and $(\widetilde M^{(T_p)-S_p}_{M_R} \otimes_e \widetilde M^{(S_p)-T_p}_{M_L})$
are {\bfseries the electric string fields\/} at the ``$ST$'', ``$MG$'' and ``$M$'' levels.
\Ee}
\vskip 11pt

\bpr This internal bilinear structure of an elementary (bisemi)fermion in three embedded (bi)shells results from section 2.7, proposition 2.5 and \cite{Pie2} and \cite{Pie4}.\epr
\vskip 11pt

\subsection{Proposition}

{\em \Bena
\item {\bfseries The energy\/} at the ``mass'' level of an elementary (bisemi)fermion is given by the sum, over all its left mass strings, of the norms of the generators of the Lie (semi)algebra $\widetilde M^{T_p-S_p}_{M_L}$.

\item {\bfseries The electric charge\/} at the ``mass'' level of an elementary (bisemi)fermion is the sum, over all sets of exchanged electric biquanta, of the pseudo-norms of the generators of the Lie (semi)algebra $\widetilde M^{S_p}_{M_L}\subset (\widetilde M^{T_p}_{M_R} \otimes_e
\widetilde M^{S_p}_{M_L})$.
\Ee}
\vskip 11pt

\bpr \Bena
\item The $(\mu ,m_\mu )$-th ``generator'' of the corresponding mass left string of the Lie (semi)algebra $\widetilde M^{T_p-S_p}_{M_L} $ is given by:
\begin{align*}
T^{T-S}_{L;M_\mu }
&= \L\{ +i\ \dfrac\hbar c\ \L(
s_{0_{L_\mu }}\ \dfrac\partial{\partial t_{0_\mu }} ,
s_{x_{L_\mu }}\ \dfrac\partial{\partial x_\mu } ,
s_{y_{L_\mu }}\ \dfrac\partial{\partial y_\mu  } ,
s_{z_{L_\mu }}\ \dfrac\partial{\partial z_\mu  } \R)\R\}\\[11pt]
&= \L\{ 
s_{0_{L_\mu }}\ m_{0_\mu } ,
s_{x_{L_\mu }}\ p_{x_\mu } ,
s_{y_{L_\mu }}\ p_{y_\mu } ,
s_{z_{L_\mu }}\ p_{z_\mu } \R\}
\end{align*}
in complete analogy with the developments of section 2.3 and \cite{Pie2} and its norm:
 \[ E_{M_{\mu ,m_\mu }}
 = \| T^{T-S}_{L;M_\mu }\|
 = (s^2_{0_{L_\mu }}\ m^2_{0_\mu }+\sum^3_{i=1} s^2_{i_{L_\mu }}\ p^2_{i_\mu })^{\half}\;, \qquad (\; x\leadsto 1\ , \; y\leadsto 2\ , \; z\leadsto 3\;),\]
 is the energy of the $(\mu ,m_\mu )$-th mass left string where
 $\vec p_\mu =\{p_{x_\mu },p_{y_\mu },p_{z_\mu }\}$ is its ``linear'' momentum and
 $\vec s_\mu =\{s_{x_\mu },s_{y_\mu },s_{z_\mu }\}$ the corresponding spin vector allowing to define the directional gradient $\vec s_\mu \centerdot \vec p_\mu $.
 
 Remark that the energy $\EE_{M_{\mu ,m_\mu }}$ results traditionally from a norm because it refers in fact to the bigenerator of the corresponding mass bistring.
 
 So, the energy at the mass level of an elementary (bisemi)fermion is given by:
 \[ \EE_M=\sum_\mu \sum_{m_\mu }\EE_{M_{\mu ,m_\mu }}\;.\]
 
 \item The $(\delta ,m_\delta )$-th generator of the corresponding set of ``$\delta $'' exchanged electric biquanta at the mass level of the Lie (semi)algebra
 $\widetilde M^{S_p}_{M_L} \subset \widetilde M^{T_p}_{M_R}
 \otimes_e \widetilde M^{S_p}_{M_L}$ is given by:
 \[\vec s_\delta \centerdot \vec p_\delta 
 =\{ 
s_{x_\delta }\centerdot p_{x_\delta },
s_{y_\delta }\centerdot p_{y_\delta },
s_{z_\delta }\centerdot p_{z_\delta }\}\]
and the corresponding electric pseudo norm will be introduced by:
\[ e_{M_{\delta ,m_\delta }}
= (s_{0_\delta }\centerdot m_{0_\delta }
\times (s_{x_\delta }\centerdot p_{x_\delta }
+s_{y_\delta }\centerdot p_{y_\delta }
+s_{z_\delta }\centerdot p_{z_\delta }))^{\half}\]
in order to take into account the electric metric \cite{Pie2}.

Then, the value of the electric charge of an elementary bisemifermion at the mass level will be given by the sum of the pseudo-norms of the electric generators of all sets of exchanged electric biquanta according to
\be e_M = \sum_\delta \sum_{m_\delta }e_{M_{\delta ,m_\delta }}\;.\tag*{\eop}
\ee
\Ee
\vskip 11pt

\subsection{Generation of bisemifermions of the 2-th and 3-th family}

\Bi
\item From section 2.2, the internal space-time structure of the elementary bisemifermions $e^-$, $u^+$ and $d^-$ of the first family was analyzed.

It is the aim of this section to recall how the internal structure of the massive leptons and quarks of the second and of the third families can be generated from the internal structure of the corresponding leptons and quarks of the first family.

\item It was seen in \cite{Pie2} that the ``$ST$'', ``$MG$'' and ``$M$'' string fields
\begin{multline*}
({}^A\widetilde M^{T-S}_{ST_R-MG_R-M_R} \otimes {}^A\widetilde M^{T-S}_{ST_L-MG_L-M_L} )\\
\equiv 
(
{}^A\widetilde M^{T-S}_{ST_R} \oplus
{}^A\widetilde M^{T-S}_{MG_R} \oplus
{}^A\widetilde M^{T-S}_{M_R}
)
\otimes_{(D)}
(
{}^A\widetilde M^{T-S}_{ST_L} \oplus
{}^A\widetilde M^{T-S}_{MG_L} \oplus
{}^A\widetilde M^{T-S}_{M_L} )
\end{multline*}
of an elementary bisemifermion ``$A$'' of the first family (i.e. $A=e^-$, $u^+$ or $d^-$), in a strong external field, could generate, by versal deformations and spreading-out of singularities of codimension 2 on all of its bisections (or bistrings), the 
``$ST$'', ``$MG$'' and ``$M$'' string fields 
$({}^B\widetilde M^{T-S}_{ST_R-MG_R-M_R} \otimes {}^B\widetilde M^{T-S}_{ST_L-MG_L-M_L} )
$ of a {\bbf bisemifermion ``$B$'' of the second family\/} (i.e. $B=\mu ^-$, $s^-$ or $c^+$) by endowing the string fields of ``$A$'' with additional string fields ``$A'$'' leading to the transformation:
\begin{multline*}
({}^{A}\widetilde M^{T-S}_{ST_R-MG_R-M_R} \otimes_{(D)} {}^A\widetilde M^{T-S}_{ST_L-MG_L-M_L} ) \cup
({}^{A'}\widetilde M^{T-S}_{ST_R-MG_R-M_R} \otimes_{(D)} {}^{A'}\widetilde M^{T-S}_{ST_L-MG_L-M_L} )\\
\To
({}^B\widetilde M^{T-S}_{ST_R-MG_R-M_R} \otimes_{(D)} {}^B\widetilde M^{T-S}_{ST_L-MG_L-M_L} )\;.\end{multline*}
By this way, the electron $e^-$ can be transformed into the muon $\mu ^-$ and the down quark $d^-$ into the strange quark $s^-$\ldots

\item If the singularities on the bisections of a bisemifermion ``$A$'' are of codimension 3, then {\bbf a bisemifermion ``$C$'' of the third family\/} can be generated by endowing the string fields of ``$A$'' with two additional string fields ``$A'$ and ``$A''$'' according to:
\begin{multline*}
({}^A\widetilde M^{T-S}_{ST_R-MG_R-M_R} \otimes_{(D)} {}^A\widetilde M^{T-S}_{ST_L-MG_L-M_L} )
\cup ({}^{A'}\widetilde M^{T-S}_{ST_R-MG_R-M_R} \otimes_{(D)} {}^{A'}\widetilde M^{T-S}_{ST_L-MG_L-M_L} )\\
\cup ({}^{A''}\widetilde M^{T-S}_{ST_R-MG_R-M_R} \otimes_{(D)} {}^{A''}\widetilde M^{T-S}_{ST_L-MG_L-M_L} )\\
\To ({}^{C}\widetilde M^{T-S}_{ST_R-MG_R-M_R} \otimes_{(D)} {}^{C}\widetilde M^{T-S}_{ST_L-MG_L-M_L} )\end{multline*}
\Ei
\vskip 11pt

\subsection[(Bisemi)photons, magnetic and electric exchange (bisemi)bosons]{(Bisemi)photons, magnetic and electric exchange (bisemi)\linebreak bosons}

\Bi
\item {\bbf A bisemiphoton at $\gamma $ biquanta\/} is a ``space'' (``$S$'') close bistring at $\gamma $ biquanta at the ``$ST$'' level $\phi ^{S_p}_{ST_R}(M_{L_{\o v_{\gamma ,m_\gamma }}}) \otimes \phi ^{S_p}_{ST_L}(M_{L_{v_{\gamma ,m_\gamma }}})$ (see proposition 2.6) endowed with their middle-ground ($MG$) and mass ($M$) covering open bistrings at $\gamma $ biquanta according to:
\begin{multline*}
\begin{aligned}[t]
(\phi ^{S_p}_{ST_R}(M^{ST}_{L_{\o v_{\gamma ,m_\gamma }}})
\oplus \phi ^{S_p}_{MG_R}(M^{MG}_{L_{\o v_{\gamma ,m_\gamma }}})
\oplus \phi ^{S_p}_{M_R}(M^{M}_{L_{\o v_{\gamma ,m_\gamma }}}))\qquad\\
\quad \otimes\
(\phi ^{S_p}_{ST_L}(M^{ST}_{L_{v_{\gamma ,m_\gamma }}})
\oplus \phi ^{S_p}_{MG_L}(M^{MG}_{L_{v_{\gamma ,m_\gamma }}})
\oplus \phi ^{S_p}_{M_L}(M^{M}_{L_{v_{\gamma ,m_\gamma }}}))
\end{aligned}\\
\begin{aligned}[t]
\simeq [
(\phi ^{S_p}_{ST_R}(M^{ST}_{L_{\o v_{\gamma ,m_\gamma }}})
\otimes_D \phi ^{S_p}_{ST_L}(M^{ST}_{L_{v_{\gamma ,m_\gamma }}}) \qquad\\
\quad \oplus\ (\phi ^{S_p}_{MG_R}(M^{MG}_{L_{\o v_{\gamma ,m_\gamma }}})
\otimes_D \phi ^{S_p}_{MG_L}(M^{MG}_{L_{v_{\gamma ,m_\gamma }}}))\\
\quad \oplus\ (\phi ^{S_p}_{M_R}(M^{M}_{L_{\o v_{\gamma ,m_\gamma }}})
\otimes_D \phi ^{S_p}_{M_L}(M^{M}_{L_{v_{\gamma ,m_\gamma }}}))]
\end{aligned}\\
\begin{aligned}[t]
\oplus\ [
(\phi ^{S_p}_{ST_R}(M^{ST}_{L_{\o v_{\gamma ,m_\gamma }}})
\otimes_m \phi ^{S_p}_{ST_L}(M^{ST}_{L_{v_{\gamma ,m_\gamma }}}) \qquad\\
\quad \oplus \ (\phi ^{S_p}_{MG_R}(M^{MG}_{L_{\o v_{\gamma ,m_\gamma }}})
\otimes_m \phi ^{S_p}_{MG_L}(M^{MG}_{L_{v_{\gamma ,m_\gamma }}}))\\
\quad \oplus (\phi ^{S_p}_{M_R}(M^{M}_{L_{\o v_{\gamma ,m_\gamma }}})
\otimes_m \phi ^{S_p}_{M_L}(M^{M}_{L_{v_{\gamma ,m_\gamma }}}))]
\end{aligned}\end{multline*}
where:
\Bean
\item $(\phi ^{S_p}_{ST_R}(M^{ST}_{L_{\o v_{\gamma ,m_\gamma }}})
\otimes_D \phi ^{S_p}_{ST_L}(M^{ST}_{L_{v_{\gamma ,m_\gamma }}})$,
$(\phi ^{S_p}_{MG_R}(M^{MG}_{L_{\o v_{\gamma ,m_\gamma }}})
\otimes_D \phi ^{S_p}_{MG_L}(M^{MG}_{L_{v_{\gamma ,m_\gamma }}}))$ and\linebreak
$(\phi ^{S_p}_{M_R}(M^{M}_{L_{\o v_{\gamma ,m_\gamma }}})
\otimes_D \phi ^{S_p}_{M_L}(M^{M}_{L_{v_{\gamma ,m_\gamma }}}))$ are respectively a close bistring on the ``$ST$'' level at ``$\gamma $'' biquanta on the bilinear algebraic conjugacy class representative
$(M^{ST}_{L_{\o v_{\gamma ,m_\gamma }}} \otimes_D M^{ST}_{L_{v_{\gamma ,m_\gamma }}})$, 
an open bistring on the ``$MG$'' level at ``$\gamma $'' biquanta on the ``$MG$'' conjugacy class representative
$(M^{MG}_{L_{\o v_{\gamma ,m_\gamma }}} \otimes_D M^{MG}_{L_{v_{\gamma ,m_\gamma }}})$, 
and an open bistring on the ``$M$'' level at $\gamma $ biquanta on the respective ``$M$' conjugacy class representative.

\item $(\phi ^{S_p}_{ST_R}(M^{ST}_{L_{\o v_{\gamma ,m_\gamma }}})
\otimes_m \phi ^{S_p}_{ST_L}(M^{ST}_{L_{v_{\gamma ,m_\gamma }}})$,
$(\phi ^{S_p}_{MG_R}(M^{MG}_{L_{\o v_{\gamma ,m_\gamma }}})
\otimes_m \phi ^{S_p}_{MG_L}(M^{MG}_{L_{v_{\gamma ,m_\gamma }}}))$ and
$ (\phi ^{S_p}_{M_R}(M^{M}_{L_{\o v_{\gamma ,m_\gamma }}})
\otimes_m \phi ^{S_p}_{M_L}(M^{M}_{L_{v_{\gamma ,m_\gamma }}}))$
are the bisections at the ``$ST$'', ``$MG$'' and ``$M$'' levels responsible for the not-simultaneous exchange of ``$\gamma $'' magnetic biquanta inside the considered bisemiphoton and {\bbf generating\/} in this manner {\bbf magnetic subfields on these levels of this bisemiphoton\/}.
\Ee

\item Let 
$\phi ^{S_p}_{ST-MG-M_R}(\o v_{\gamma ,m_\gamma })
\otimes \phi ^{S_p}_{ST-MG-M_L}(v_{\gamma ,m_\gamma })$ denote in condensed form the ``$ST$'', ``$MG$'' and ``$M$'' structures of the considered bisemiphoton at ``$\gamma $'' biquanta.

Then, this bisemiphoton can join another bisemiphoton at ``$\delta $'' biquanta, $\delta \ge\gamma $ or $\delta \le\gamma $, according to the map:
\begin{multline*}
\Ds^{[\gamma ]\to[\gamma +\delta ]}\RL: \quad
\phi ^{S_p}_{ST-MG-M_R}(\o v_{\gamma ,m_\gamma })
\otimes \phi ^{S_p}_{ST-MG-M_L}(v_{\gamma ,m_\gamma })\\
\To \phi ^{S_p}_{ST-MG-M_R}(\o v_{\gamma +\delta ,m_{\gamma +\delta }})
\otimes \phi ^{S_p}_{ST-MG-M_L}(v_{\gamma+\delta  ,m_{\gamma+\delta } })
\end{multline*}
which is {\bbf a deformation\/}, in the sense of Mazur \cite{Maz}, \cite{Rib}, of a Galois subrepresentation {\bbf corresponding to an equivalence class of lift\/} at the ``$ST$'', ``$MG$'' and ``$M$'' levels sending the bisemiphoton at ``$\gamma $'' biquanta into a bisemiphoton at ``$(\gamma +\delta )$'' biquanta.

\item The inverse transformation can also be envisaged in such a way that a bisemiphoton at $(\gamma +\delta )$ biquanta can split into two bisemiphotons at ``$\gamma $'' and ``$\delta $'' biquanta.

The frame of {\bf the Bose-Einstein statistics\/} is then reached for (bisemi)photons since they can join together ``on the same level''.

And, {\bf the deformations $\Ds^{[\gamma ]\to[\gamma +\delta ]}\RL$ and their inverses
$\Ds^{{}^{-1}[\gamma ]\to[\gamma +\delta ]}\RL$ correspond to the raising and lowering operators\/} of quantum field theories {\bf leading to quantization rules\/}.

\item Remark that {\bbf a set of ``$\beta $'' magnetic biquanta\/}, $\beta \le\gamma $, at the ``$ST$'', ``$MG$'' and ``$M$'' levels, {\bbf is a (bisemi)boson\/} since these exchanged magnetic biquanta can joint ``$(\gamma -\beta )$'' magnetic biquanta: they thus obey the Bose-Einstein statistics.

\item Similarly, a set of exchanged electric biquanta is an electric (bisemi)boson.
\Ei
\vskip 11pt

\subsection{Proposition}

{\em \Bena
\item The (bisemi)photons and the sets of magnetic and electric exchange biquanta are (bisemi)bosons, obeying the Bose-Einstein statistics.

\item The elementary (bisemi)fermions obey the Fermi-Dirac statistics.
\Ee}
\vskip 11pt

\bpr
\Bi
\item Part 1) was already proved in section 2.11.

\item Part 2) results simply from the fact that two bisemifermions at ``ordinary'' energies (for example two (bisemi)electrons) cannot amalgamate because {\bbf there is an obstruction due to their electric charges\/} generating an electric field between their \lr time string semifields and their \rl ``space'' string semifields throughout the emergence point, i.e. the origin of the considered bisemifermion.\epr
\Ei
\vskip 11pt

\subsection{Introducing composite bisemiparticles}

Since the beginning of this chapter, the space-time internal structure of the elementary bisemifermions and bisemibosons has been reviewed.  It is now the aim of the rest of this chapter to recall the internal structure of composite bisemifermions, i.e. (bisemi)baryons, and composite ``strong'' bisemibosons, i.e. mesons, as developed in \cite{Pie2}.
\vskip 11pt

\subsection{The internal structure of bisemibaryons}

\Bi
\item In \cite{Pie2}, the ``time'' string semifield at the ``$ST$'' level of a \lr semibaryon was introduced by the existence of {\bbf a ``core'' time semifield $\widetilde M^{({\rm Bar});T}_{ST_L}$ \resp{$\widetilde M^{({\rm Bar});T}_{ST_R}$}} from which the ``time'' string semifields $\widetilde M^{q_i;T}_{ST_L}$ \resp{$\widetilde M^{q_i;T}_{ST_R}$}, $1\le i\le 3$, of the ``$ST$'' structures of the three \lr semiquarks are generated by the algebraic smooth endomorphism $\EE_{t_L}$ \resp{$\EE_{t_R}$} according to:
\begin{align*}
\EE_{t_L} : \quad \widetilde M^{({\rm Bar});T}_{ST_L} &\To
\widetilde M^{*({\rm Bar});T}_{ST_L}\bigoplus^3_{i=1} \widetilde M^{q_i;T}_{ST_L}\\[11pt]
\rresp{\EE_{t_R} : \quad \widetilde M^{({\rm Bar});T}_{ST_R} &\To
\widetilde M^{*({\rm Bar});T}_{ST_R}\bigoplus^3_{i=1} \widetilde M^{q_i;T}_{ST_R}}
\end{align*}
in such a way that the time string semifields of the three semiquarks be connected to the core time semifield of the semibaryon.

By this way, {\bbf the confinement \cite{G-W} of the three (semi)quarks\/} inside a (semi)baryon finds a natural explanation \cite{Kok}.

\item As in section 2.2, the ``space'' semifields
$\widetilde M^{q_i;S}_{ST_L}$ \resp{$\widetilde M^{q_i;S}_{ST_R}$} corresponding to the ``time'' semifields of the three semiquarks are generated by the
$(\gamma _{t_{i_L}}\circ \EE_{i_L})$
\resp{$(\gamma _{t_{i_R}}\circ \EE_{i_R})$} morphisms, leading to {\bbf the ``space-time'' semifields\/}
\begin{align*}
 \widetilde M^{({\rm Bar});T-S}_{ST_L} &=
\widetilde M^{*({\rm Bar});T}_{ST_L}\bigoplus^3_{i=1} \widetilde M^{q_i;T-S}_{ST_L}\\[11pt]
\rresp{\widetilde M^{({\rm Bar});T-S}_{ST_R} &=
\widetilde M^{*({\rm Bar});T}_{ST_R}\bigoplus^3_{i=1} \widetilde M^{q_i;T-S}_{ST_R}}
\end{align*}
{\bbf at the ``$ST$'' level of a \lr semibaryon\/}.

\item And, {\bbf the space-time fields at the ``$ST$'' level of a bisemibaryon are\/}:
\[  \widetilde M^{({\rm Bar});T-S}_{ST\RL} =
\widetilde M^{({\rm Bar});T-S}_{ST_R}
\otimes \widetilde M^{({\rm Bar});T-S}_{ST_L}\;.\]

\item Let
\begin{multline*}
DT^{({\rm Bar})}_L 
= \L\{ i\hbar_{ST}\ s_0\ dt_{C_0},
\L\{ i\hbar_{ST}\ G^{-1}(\rho )\ s^{(1)}_{C_0}\ dt_{0},
i\ \dfrac{\hbar_{ST}}{C_{t\to r;ST}}\ G^{-1}(\rho )\ s^{(1)}_{x_L}\ dx,\R.\R.\\[11pt]
\L.\dots,
i\ \dfrac{\hbar_{ST}}{C_{t\to r;ST}}\ G^{-1}(\rho )\ s^{(1)}_{z_L}\ dz\R\},
\L\{\phantom{\dfrac12}\dots\phantom{\dfrac12}\R\}, \\[11pt]
\L.\L\{ i\hbar_{ST}\ G^{-1}(\rho )\ s^{(3)}_{0_L}\ dt_{0}, \dots,
i\ \dfrac{\hbar_{ST}}{C_{t\to r;ST}}\ G^{-1}(\rho )\ s^{(3)}_{z_L}\ dz\R\}\R\}
\end{multline*}
be the left differential operator acting on 
$\widetilde M^{({\rm Bar});T-S}_{ST_L}$ and let
$DT^{({\rm Bar})}_R$ be its right correspondent,

where:
\Bi
\item $G(\rho )$ is the defined strong constant in the frame of AQT (see \cite{Pie2}),
\item $\hbar_{ST}$ and $C_{t\to r;ST}$ were introduced in section 2.3,
\item the indices $(1)$, $(2)$ and $(3)$ refer to the semiquarks.
\Ei

Then, the {\bbf operator-valued string fields of the $ST$-level of the\/} internal vacuum of the considered {\bbf bisemibaryon\/} result from the morphism:
\[ DT^{({\rm Bar})}_R\otimes DT^{({\rm Bar})}_L : \quad
\widetilde M^{({\rm Bar});T-S}_{ST_R} \otimes \widetilde M^{({\rm Bar});T-S}_{ST_L}
\To
\widetilde M^{({\rm Bar});T_p-S_p}_{ST_R}\otimes \widetilde M^{({\rm Bar});T_p-S_p}_{ST_L}
\]
where $(\widetilde M^{({\rm Bar});T_p-S_p}_{ST_R}\otimes \widetilde M^{({\rm Bar});T_p-S_p}_{ST_L}$ is a perverse bisemisheaf of which function consists in rotating its bisections or bistrings.

\item Under versal deformations and blowups of these, the ``$ST$'' level fields of the bisemibaryon can join {\bbf the two successive covering ``$MG$'' and ``$M$'' fields\/} in such a way that we have the embeddings:
\[
\widetilde M^{({\rm Bar});T_p-S_p}_{ST_R}
\otimes \widetilde M^{({\rm Bar});T_p-S_p}_{ST_L}
\subset \widetilde M^{({\rm Bar});T_p-S_p}_{MG_R}
\otimes \widetilde M^{({\rm Bar});T_p-S_p}_{MG_L}
\subset \widetilde M^{({\rm Bar});T_p-S_p}_{M_R}
\otimes \widetilde M^{({\rm Bar});T_p-S_p}_{M_L}\;.\]

\item Referring to section 2.3 and \cite{Pie2}, the mass operator
$\MM^{({\rm Bar})}_L$ of a left semibaryon is given by:
\begin{multline*}
\MM^{({\rm Bar})}_L 
= \L\{ +i\hbar\ s_0\ \dfrac\partial{\partial t_{C_0}},
\L\{ i\hbar\ G^{-1}(\rho )\ s^{(1)}_{0_L}\ \dfrac\partial{\partial t_{0}},
i\ \dfrac{\hbar}{C_{t\to r;M}}\ G^{-1}(\rho )\ s^{(1)}_{x_L}\ \dfrac\partial{\partial x},\R.\R.\\[11pt]
\L.\dots,
+i\ \dfrac{\hbar}{C_{t\to r;M}}\ G^{-1}(\rho )\ s^{(1)}_{z_L}\ \dfrac\partial{\partial z}\R\},
\L\{\phantom{\dfrac12}\dots\phantom{\dfrac12}\R\}, \L\{\phantom{\dfrac12}\dots\phantom{\dfrac12}\R\},\\[11pt]
\L.\L.
i\ \dfrac{\hbar}{C_{t\to r;M}}\ G^{-1}(\rho )\ s^{(3)}_{z_L}\ \dfrac\partial{\partial z}\R\}\R\}
\end{multline*}
\Ei
\vskip 11pt

\subsection{Proposition}

{\em
At the mass level ``$M$'', {\bfseries the\/} (operator-valued) {\bfseries string fields of a (bisemi)baryon are given by\/}:
\begin{multline*}
\widetilde M^{({\rm Bar});T_p-S_p}_{M_R}
\otimes \widetilde M^{({\rm Bar});T_p-S_p}_{M_L}
= 
(\widetilde M^{*({\rm Bar});T_p}_{M_R} \otimes\widetilde M^{*({\rm Bar});T_p}_{M_L})\\[11pt]
\bigoplus^3_{i=1}
(\widetilde M^{q_i;T_p-S_p}_{M_R} \otimes\widetilde M^{q_i;T_p-S_p}_{M_L})
\bigoplus^3_{\begin{subarray}{c}i=j=1\\i\neq j\end{subarray}}
(\widetilde M^{q_i;T_p-S_p}_{M_R} \otimes\widetilde M^{q_j;T_p-S_p}_{M_L})\\[11pt]
\bigoplus^3_{i=1}
(\widetilde M^{*({\rm Bar});T_p}_{M_R} \otimes\widetilde M^{q_i;T_p-S_p}_{M_L})
\bigoplus^3_{i=1}
(\widetilde M^{q_i;T_p-S_p}_{M_R}  \otimes\widetilde M^{*({\rm Bar});T_p}_{M_L})
\end{multline*}
where:
\Bean
\item $(\widetilde M^{*({\rm Bar});T_p}_{M_R} \otimes\widetilde M^{*({\rm Bar});T_p}_{M_L})$ is the {\bfseries  core central ``time'' string field of the bisemibaryon\/},

\item $(\widetilde M^{q_i;T_p-S_p}_{M_R} \otimes\widetilde M^{q_i;T_p-S_p}_{M_L})
= (\widetilde M^{q_i;T_p-S_p}_{M_R} \otimes_D\widetilde M^{q_i;T_p-S_p}_{M_L})
\oplus (\widetilde M^{q_i;S_p}_{M_R} \otimes_m\widetilde M^{q_i;S_p}_{M_L})
\oplus (\widetilde M^{q_i;(T_p)-S_p}_{M_R} \otimes_e\widetilde M^{q_i;(S_p)-T_p}_{M_L})$, $1\le i\le 3$, is the {\bfseries\boldmath ``mass'' string field of the $i$-th bisemiquark\/} decomposing into the diagonal space-time string field
$(\widetilde M^{q_i;T_p-S_p}_{M_R} \otimes\widetilde M^{q_i;T_p-S_p}_{M_L})$,
the magnetic string field
$(\widetilde M^{q_i;S_p}_{M_R} \otimes_m\widetilde M^{q_i;S_p}_{M_L})$
and into the electric string field
$(\widetilde M^{q_i;(T_p)-S_p}_{M_R} \otimes_e\widetilde M^{q_i;(S_p)-T_p}_{M_L})$
responsible for the (bisemi)quark electric charge having an absolute value $\L| \F13\R|\ e$ or 
$\L| \F23\R|\ e$.

\item $(\widetilde M^{q_i;T_p-S_p}_{M_R} \otimes\widetilde M^{q_j;T_p-S_p}_{M_L})
=(\widetilde M^{q_i;T_p-S_p}_{M_R} \otimes_D\widetilde M^{q_j;T_p-S_p}_{M_L})
\oplus(\widetilde M^{q_i;S_p}_{M_R} \otimes_m\widetilde M^{q_j;S_p}_{M_L})
\oplus(\widetilde M^{q_i;(T_p)-S_p}_{M_R} \otimes_e\widetilde M^{q_j;(S_p)-T_p}_{M_L})$
is the {\bfseries\boldmath mixed string field of interaction between the $i$-th right semiquark and the $j$-th left semiquark\/}: it decomposes into:
\Bi
\item the diagonal space-time string field
$(\widetilde M^{q_i;T_p-S_p}_{M_R} \otimes_D\widetilde M^{q_j;T_p-S_p}_{M_L})$ which is of gravitational nature responsible for the exchange of gravitational biquanta,

\item the magnetic string field of interaction
$(\widetilde M^{q_i;S_p}_{M_R} \otimes_m\widetilde M^{q_j;S_p}_{M_L})$,

\item the electric string field of interaction
$(\widetilde M^{q_i;(T_p)-S_p}_{M_R} \otimes_e\widetilde M^{q_j;(S_p)-T_p}_{M_L})$.
\Ei

\item $(\widetilde M^{*({\rm Bar});T_p}_{M_R} \otimes\widetilde M^{q_i;T_p-S_p}_{M_L})
=(\widetilde M^{*({\rm Bar});T_p}_{M_R} \otimes\widetilde M^{q_i;T_p}_{M_L})
\oplus(\widetilde M^{*({\rm Bar});T_p}_{M_R} \otimes_e\widetilde M^{q_i;S_p}_{M_L})$
and 

$(\widetilde M^{q_i;T_p-S_p}_{M_R}  \otimes\widetilde M^{*({\rm Bar});T_p}_{M_L})
=(\widetilde M^{q_i;T_p}_{M_R}  \otimes\widetilde M^{*({\rm Bar});T_p}_{M_L})
\oplus (\widetilde M^{q_i;S_p}_{M_R}  \otimes_e\widetilde M^{*({\rm Bar});T_p}_{M_L})$

are respectively the {\bfseries mixed ``strong'' string fields of interaction\/} between the core central ``time'' string semifield of the right semibaryon and the space-time string semifield of the $i$-th left semiquark and between the space-time string semifield of the $i$-th right semiquark and the core central ``time'' string semifield of the left semibaryon.
\Ee

\Bi
\item $(\widetilde M^{*({\rm Bar});T_p}_{M_R} \otimes\widetilde M^{q_i;T_p-S_p}_{M_L})$
decomposes into {\bfseries a mixed ``strong'' time string field 
$(\widetilde M^{*({\rm Bar});T_p}_{M_R} \otimes\widetilde M^{q_i;T_p}_{M_L})$
of gravitational nature\/} and into {\bfseries a mixed ``strong'' electric string field\/}
$(\widetilde M^{*({\rm Bar});T_p}_{M_R} \otimes_e\widetilde M^{q_i;S_p}_{M_L})$.

\item The direct sum
$[(\widetilde M^{*({\rm Bar});T_p}_{M_R} \otimes_e\widetilde M^{q_i;S_p}_{M_L})
\oplus (\widetilde M^{q_j;S_p}_{M_R} \otimes_e\widetilde M^{*{\rm Bar});T_p}_{M_L})]$ of the mixed electric ``strong'' string fields, as well as the direct sum
$[(\widetilde M^{*({\rm Bar});T_p}_{M_R} \otimes_e\widetilde M^{q_i;T_p}_{M_L})
\oplus (\widetilde M^{q_j;T_p}_{M_R} \otimes\widetilde M^{*{\rm Bar});T_p}_{M_L})]$ 
of the mixed gravitational ``strong'' string fields are probably responsible for {\bfseries the generation of mesons\/} of quark composition $\o q_j\ q_i$.
\Ei
}
\vskip 11pt

\bpr
\Bean
\item The ``mass'' core central ``time'' string field
$(\widetilde M^{*({\rm Bar});T_p}_{M_R} \otimes\widetilde M^{*{\rm Bar});T_p}_{M_L})$ 
is characterized by a number on $n_B$ bistrings having generated by the smooth biendmorphism $(\EE_{t_R}\otimes \EE_{t_L})$ the three sets of ``time'' bistrings of the mass fields of the three bisemiquarks according to section 2.14.

\item The mass string field 
$(\widetilde M^{q_i;T_p-S_p}_{M_R})
\otimes \widetilde M^{q_i;T_p-S_p}_{M_L})$
of the $i$-th bisemiquark decomposes, as developed in proposition 2.8, into a diagonal space-time string field, a magnetic string field and an electric string field responsible for the value $\L| \F13\R|\ e$ or 
$\L| \F23\R|\ e$ of the electric charge in such a way that the electric charge of the considered (bisemi)baryon takes an integer value.

\item The diagonal space-time string field
$(\widetilde M^{q_i;T_p-S_p}_{M_R})
\otimes_D \widetilde M^{q_j;T_p-S_p}_{M_L})$ of the mixed string field of  interaction between the $i$-th right semiquark and the $j$-th left semiquark is of gravitational nature as proved in \cite{Pie2} because the respective ``mass'' bioperator can be considered as an operator of mixed acceleration.

\item {\bf The mixed strong string fields\/}
$(\widetilde M^{*({\rm Bar});T_p}_{M_R} \otimes\widetilde M^{q_i;T_p-S_p}_{M_L})$ and
$(\widetilde M^{q_i;T_p-S_p}_{M_R} \otimes\widetilde M^{*({\rm Bar});T_p}_{M_L})$, $1\le i\le 3$,  {\bf are responsible for the strong force\/} inside the considered bisemibaryon in such a way that the bisemiquarks be steadily linked to the core central ``time'' string field leading to their confinement.

Under some external strong perturbation, these mixed strong string fields are able to generate massive mesons according to the following procedure:

\Bi
\item Let $(\widetilde M^{*({\rm Bar});T_p}_{M_R} \otimes\widetilde M^{q_i;T_p-S_p}_{M_L})\oplus 
(\widetilde M^{q_j;T_p-S_p}_{M_R} \otimes\widetilde M^{*({\rm Bar});T_p}_{M_L})$
be the mixed strong string fields between:
\Be
\item the right central core semifield of the right semibaryon and the ``space-time'' mass semifield of the $i$-th left semiquark,
\item the ``space-time'' mass semifield of the $j$-th right semiquark and the left central core semifield of the left semibaryon.
\Ee

\item Assume that the set of ``exchanged'' \rl ``time'' strings of 
$(\widetilde M^{*({\rm Bar});T_p}_{M_R} $ \resp{$\widetilde M^{*({\rm Bar});T_p}_{M_L})$} generates by 
$(\gamma _{t_R\to r_R}\circ \EE_R)$
\resp{$(\gamma _{t_L\to r_L}\circ \EE_L)$} morphisms their corresponding space strings.

\item So, we get a set of ``~$\beta $~'', $\beta \in\NN$, exchanged right (resp. ``~ $\gamma $~'' exchanged left) space-time strings of
$(\widetilde M^{*({\rm Bar});T_p-S_p}_{M_R} $ \resp{$\widetilde M^{*({\rm Bar});T_p-S_p}_{M_L})$}
which are assumed to ``join'' a corresponding set of ``$\beta $'' exchanged left
 (resp. ``~ $\gamma $~'' exchanged right) space-time strings of
$M^{q_i;T_p-S_p}_{M_L})$ \resp{$M^{q_j;T_p-S_p}_{M_R})$} giving then the direct sum of semifields:
\[ (\widetilde M^{*({\rm Bar});T_p-S_p}_{M_R}\{\beta \} \otimes\widetilde M^{q_i;T_p-S_p}_{M_L}\{\beta \})
\oplus
(\widetilde  M^{q_j;T_p-S_p}_{M_L})\{\gamma \} \otimes \widetilde M^{*({\rm Bar});T_p-S_p}_{M_L}\{\gamma  \} )\;.\]
Each tensor product is assumed to split into a diagonal, magnetic and electric tensor product responsible respectively for diagonal bistrings, magnetic exchanged bistrings and electric exchanged bistrings.

\item The mixed interaction strong field of strings
$(\widetilde M^{*({\rm Bar});T_p-S_p}_{M_R}\{\beta \} \otimes\widetilde M^{q_i;T_p-S_p}_{M_L}\{\beta \})$ thus has the structure of a bisemiquark $q_i$ at the mass level and the other mixed interaction strong field of strings
$(\widetilde  M^{q_j;T_p-S_p}_{M_L})\{\gamma \} \otimes \widetilde M^{*({\rm Bar});T_p-S_p}_{M_L}\{\gamma  \} )$ has the structure of a bisemiquark $\o q_j$  in such a way that their direct sum $\o q_j\oplus q_i$, written in condensed form $\o q_j\ q_i$, is a meson with bisemiquark structure $\o q_j\ q_i$.

\item Remark that this meson $\o q_j\ q_i$ will be of ``scalar'' nature if the strings of
$\widetilde M^{q_i;T_p-S_p}_{M_L}\{\beta \}$ rotate in the opposite sense of those of
$\widetilde M^{*({\rm Bar});T_p-S_p}_{M_L}\{\gamma  \}$.

On the other hand, $\o q_j\ q_i$ will be of ``vectorial'' nature if the strings of
$\widetilde M^{q_i;T_p-S_p}_{M_L}\{\beta \}$ rotate in the same sense of
$\widetilde M^{*({\rm Bar});T_p-S_p}_{M_L}\{\gamma  \}$.\epr
\Ei\Ee
\vskip 11pt

\subsection{Corollary}

{\em
At the ``$ST$'', ``$MG$'' and ``$M$'' levels, {\bfseries a right and a left semibaryon of a given bisemibaryon interact by means of\/}:
\Bean
\item the electric charges and the magnetic moments of the 3 (bisemi)quarks.

\item a gravito-electro-magnetic field resulting from the bilinear interactions between the right and the left semiquarks of different bisemiquarks.

\item a strong gravitational and electric field resulting from the bilinear interactions between the central core structures of the left and right semibaryons and, respectively, the right and left semiquarks.
\Ee}

\section{Gravito-electro-magnetic fields of interaction}

In this chapter, the interactions between a set of bisemiparticles are taken up from two points of view:
\BeAn
\item {\bf The first approach is explicit in the sense that the interactions\/} between bisemiparticles {\bf are considered in the bilinear frame of the global program of Langlands\/} based on the reducible representation (space) $\Repsp \GL_{2J}(L_{\o V}\times L_V)$ of the general algebraic bilinear semigroup
$\GL_{2J}(L_{\o V}\times L_V)$ of order $2J$~.

This allows to describe explicitly the gravito-electro-magnetic fields of interaction between ``~$J$~'' interacting bisemiparticles.
\vskip 11pt

\item {\bf The second approach\/}, which is {\bf more explicit, is based on the study of biconnections\/} leading to Maxwell equations in a more general bilinear mathematical frame which allows to introduce in these the gravitational field(s): this approach was extensively developed in the algebraic quantum theory (AQT) \cite{Pie2} to which we refer.
\Ee
\vskip 11pt

\subsection*{A) Explicit description of the interaction fields of a set of interacting bisemiparticles}

\subsection[The (operator valued) string fields of a set of ``~$J$~'' interacting bisemiparticles]{\bbf The (operator valued) string fields of a set of ``~$J$~'' interacting bisemiparticles}

\Bi
\item Assume that we have a set of ``~$J$~'' interacting bisemiparticles.  The ``time'' field of their ``~$ST$~'' level is given by a bisemisheaf of differentiable functions on the non-orthogonal completely reducible representation space $\Repsp \GL_{2J}(L_{\o V}\times L_V)$ of the   bilinear general semigroup
$\GL_{2J}(L_{\o V}\times L_V)$ where $L_V$ \resp{$L_{\o V}$} is the sum of the \lr completions referring to the considered \lr semiparticles.
\vskip 11pt

\item According to \cite{Pie3}, {\bbf the completely reducible non-orthogonal representation space of $\GL_{2J}(L_{\o V}\times L_V)$ decomposes as follows\/}:
\begin{align*}
&\Repsp(\GL_{2J}(L_{\o V}\times L_V))
\equiv \Repsp \GL_{2J}\L(\L(\sum\limits^J_{i=1}L_{\o v_i}\R)\times \L(\sum\limits^J_{i=1}L_{v_j}\R)\R)\\[11pt]
&\qquad \simeq 
\Repsp (\GL_{2_1}(L_{\o v_1}\times L_{v_1}))
\begin{aligned}[t]
&\times \dots\times
\Repsp (\GL_{2_i}(L_{\o v_i}\times L_{v_i}))\\
&\quad \times \dots\times
\Repsp (\GL_{2_J}(L_{\o v_J}\times L_{v_J}))\end{aligned}
\\\noalign{\newpage}
&\qquad \simeq 
\L( \Boxplus^J_{i=1}
\Repsp (T^t_{2_i}(L_{\o v_i}))\R) 
\otimes 
\L( \Boxplus^J_{j=1}
\Repsp (T_{2_j}(L_{v_j}))\R)
\\[11pt]
&\qquad = \begin{aligned}[t]
&\Boxplus^J_{i=1}\L(
\Repsp (T^t_{2_i}(L_{\o v_i}))\otimes 
\Repsp (T_{2_i}(L_{v_i}))\R) \\
&\quad \Boxplus^J_{i\neq j=1}\L(
\Repsp (T^t_{2_i}(L_{\o v_i}))\otimes 
\Repsp (T_{2_j}(L_{v_j}))\R) \end{aligned}
\\[11pt]
&\qquad = 
\Boxplus^J_{i=1}
\Repsp (\GL_{2_i}(L_{\o v_i}\times L_{v_i}))
\Boxplus^J_{i\neq j=1}
\Repsp (T^t_{2_i}((L_{\o v_i})\times T_{2j}(L_{v_j})))
\end{align*}
in such a way that:
\Bena
\item $M^{T}_{ST_R}(i)\otimes M^{T}_{ST_L}(i)
\equiv 
\Repsp (\GL_{2_i}(L_{\o v_i}\times L_{v_i}))$ is the ``time'' field (i.e. structure) of the ``~$ST$~'' level of the $i$-th bisemiparticle;

\item $M^{T}_{ST_R}(i)\otimes M^{T}_{ST_L}(j)
\equiv 
\Repsp (T^t_{2_i}(L_{\o v_i})\times T_{2_j}(L_{v_j}))$ is the ``time'' interaction field(s) at the ``~$ST$~'' level between the $i$-th right semiparticle and the $j$-th left semiparticle;
\Ee
if we take into account the {\bf Gauss bilinear decomposition\/}:
\[ \GL_{2_i}(L_{\o v_i}\times L_{v_i})
= T^t_{2_i}(L_{\o v_i})\times T_{2_i}(L_{v_i})\]
of the bilinear algebraic semigroup of order 2 over the product $(L_{\o v_i}\times L_{v_i})$ of the sets of completions:
\begin{align*}
L_{v_i} &= \{ L_{v_{i_1}},\dots,L_{v_{i_{\mu ,m_\mu }}},\dots,L_{v_{i_{q,m_q}}}\}\\
\text{and} \quad 
L_{\o v_i} &= \{ L_{\o v_{i_1}},\dots,L_{\o v_{i_{\mu ,m_\mu }}},\dots,L_{\o v_{i_{q,m_q}}}\}\;.\end{align*}
\vskip 11pt

\item Referring to section 2.2, the $(\gamma _{t_{R\times L}\to r_{R\times L}}\circ E\RL)$ morphisms transform the ``time'' fields into reduced 
``time'' fields and complementary ``space'' fields according to:
\[\gamma ^{(i)}_{t_{R\times L}\to r_{R\times L}}\circ E^{(i)}\RL:
\quad M^{T}_{ST_R}(i)\otimes M^{T}_{ST_L}(i)
\To M^{T-S}_{ST_R}(i)\otimes M^{T-S}_{ST_L}(i)
\;.\]
\vskip 11pt

\item {\bbf The (operator-valued) string field at the ``~$ST$~'' level\/} corresponding to the ``space-time'' structure $(M^{T-S}_{ST_R}(i)\otimes M^{T-S}_{ST_L}(i))$ {\bbf of the $i$-th bisemiparticle is the bisemisheaf
$(\widetilde M^{T_p-S_{p}}_{ST_R}(i)\otimes \widetilde M^{T_p-S_{p}}_{ST_L}(i))$
 of differentiable functions on\/}
 $(M^{T-S}_{ST_R}(i)\otimes M^{T-S}_{ST_L}(i))$~.
 \Ei
 \vskip 11pt

\subsection{Proposition}

{\em The (operator-valued) string fields at the ``~$ST$~'' level of a set of ``~$J$~'' interacting bisemiparticles are given by:
\[
(\widetilde M^{T_p-S_p}_{ST_R}(J)\otimes
\widetilde M^{T_p-S_p)}_{ST_L}(J))
= \bigoplus^J_{i=1}(\widetilde M^{T_p-S_{p}}_{ST_R}(i)\otimes
\widetilde M^{T_p-S_{p}}_{ST_L}(i))
\bigoplus^J_{i\neq j=1}(\widetilde M^{T_p-S_{p}}_{ST_R}(i)\otimes
\widetilde M^{T_p-S_{p}}_{ST_L}(j))
\]
where:
\Be
\item the direct sum $\bigoplus^J_{i=1}(\widetilde M^{T_p-S_{p}}_{ST_R}(i)\otimes
\widetilde M^{T_p-S_{p}}_{ST_L}(i))$ is the structure field at the ``~$ST$~'' level of a set of $J$ non-interacting (i.e. free) (bisemi)particles verifying the conditions of non connectivity between the $i$-th and $j$-th bisemisheaves:
\[ (\widetilde M^{T_p-S_{p }}_{ST_R}(i)\otimes
\widetilde M^{T_p-S_{p }}_{ST_L}(i))
\cap (\widetilde M^{T_p-S_{p}}_{ST_R}(j)\otimes
\widetilde M^{T_p-S_{p}}_{ST_L}(j))=\emptyset\;;\]

\item the mixed direct sum refers to the bilinear interaction fields at the ``~$ST$~'' level between the right and left semiparticles belonging to different bisemiparticles.
\Ee
}
\vskip 11pt

\bpr The assertions of this proposition are a direct consequence of the developments of section 3.1.\epr
\vskip 11pt

\subsection[``~$ST$~'', ``~$MG$~'' and ``~$M$~'' levels of ``~$J$~'' interacting bisemiparticles]{\bbf ``~$ST$~'', ``~$MG$~'' and ``~$M$~'' levels of ``~$J$~'' interacting bisemiparticles}

\Bi
\item The (operator-valued) string fields at the ``~$ST$~'' level of a set of  ``~$J$~'' interacting bisemiparticles can join, by versal deformations and blowups of these, the two successive covering  ``~$MG$~'' and ``~$M$~'' string fields
$(\widetilde M^{T_p-S_ p}_{MG_R}(J)\otimes
\widetilde M^{T_p-S_ p}_{MG_L}(J))$ and
$(\widetilde M^{T_p-S_ p}_{M_R}(J)\otimes
\widetilde M^{T_p-S_ p }_{M_L}(J))$~.

The (operator-valued) string fields at the ``~$ST$~'', ``~$MG$~'' and ``~$M$~'' levels for these ``~$J$~'' interacting bisemiparticles will be written in condensed form
:
\[
(\widetilde M^{T_p-S_ p }_{ST_R-MG_R-M_R}(J)\otimes
\widetilde M^{T_p-S_ p }_{ST_L-MG_L-M_L}(J))
\quad \text{or} \quad
(\widetilde M^{T_p-S_ p }_{V-M_R}(J)\otimes
\widetilde M^{T_p-S_ p }_{V-M_L}(J))\]
where $V-M_R$ \resp{$V-M_L$} means \rl internal vacuum and mass structure(s).
\vskip 11pt

\item It is assumed that the interactions between the ``~$ST$~'', ``~$MG$~'' and ``~$M$~''  fields are, in first approximations, negligible (see \cite{Pie2}): so,
$(\widetilde M^{T_p-S_ p }_{V-M_R}(J)\otimes
\widetilde M^{T_p-S_ p }_{V-M_L}(J))$ develops according to:
\[
(\widetilde M^{T_p-S_ p}_{V-M_R}(J)\otimes
\widetilde M^{T_p-S_ p }_{V-M_L}(J))
\begin{aligned}[t]
&  =(\widetilde M^{T_p-S_ p }_{ST_R}(J)\oplus
\widetilde M^{T_p-S_ p }_{MG_R}(J)\oplus
\widetilde M^{T_p-S_ p}_{M_R}(J))\\
& \qquad \otimes
(\widetilde M^{T_p-S_ p }_{ST_L}(J)
 \oplus
\widetilde M^{T_p-S_ p }_{MG_L}(J)\oplus
\widetilde M^{T_p-S_ p }_{M_L}(J))
\\
& \simeq (\widetilde M^{T_p-S_ p }_{ST_R}(J)\otimes
\widetilde M^{T_p-S_ p }_{ST_L}(J))\oplus
(\widetilde M^{T_p-S_ p }_{MG_R}(J)\\
& \qquad \otimes
\widetilde M^{T_p-S_ p }_{MG_L}(J))
\oplus
(\widetilde M^{T_p-S_ p }_{M_R}(J)\otimes
\widetilde M^{T_p-S_ p }_{M_L}(J))\;.
\end{aligned}
\]
\vskip 11pt

\item The following developments concerning the interaction fields of this set of ``~$J$~'' interacting bisemiparticles will be made on these three embedded levels ``~$ST$~'', ``~$MG$~'' and ``~$M$~''.
\Ei
\vskip 11pt

\subsection{Interaction fields of interacting bisemileptons}

{\bbf The (operator-valued) string fields at the ``~$ST$~'', ``~$MG$~'' and ``~$M$~'' levels of a set of ``~$J$~'' interacting bisemileptons are given by:}
\begin{multline*}
\qquad (\widetilde M^{T_p-S_ p}_{V-M_R}(\ell_J)
\otimes \widetilde M^{T_p-S_ p}_{V-M_L}(\ell_J))\\
= \bigoplus^J_{i=1}(\widetilde M^{T_p-S_ p}_{V-M_R}(\ell_i)
\otimes \widetilde M^{T_p-S_ p}_{V-M_L}(\ell_i))
 \bigoplus^J_{i\neq j=1}(\widetilde M^{T_p-S_ p}_{V-M_R}(\ell_i)
\otimes \widetilde M^{T_p-S_ p}_{V-M_L}(\ell_j))
\end{multline*}
where:
\Bi
\item $\ell_i,\ell_j$ (and $\ell_J$~) are the indices for the left and right semileptons.
\item $(\widetilde M^{T_p-S_ p}_{V-M_R}(\ell_i)
\otimes \widetilde M^{T_p-S_ p}_{V-M_L}(\ell_i))$ are the ``~$ST$~'', ``~$MG$~'' and ``~$M$~'' internal string fields of the $i$-th bisemilepton, decomposing into diagonal internal fields (~$\pt\otimes_D\pt$~), magnetic internal fields (~$\pt\otimes_m\pt$~) and electric internal fields at these 3 levels according to proposition 2.5.

\item $
(\widetilde M^{T_p-S_ p}_{V-M_R}(\ell_i)
\otimes \widetilde M^{T_p-S_ p}_{V-M_L}(\ell_j))$

\quad $\begin{aligned}[t]
&=(\widetilde M^{T_p-S_ p}_{V-M_R}(\ell_i)
\otimes_D \widetilde M^{T_p-S_ p}_{V-M_L}(\ell_j))
 \oplus (\widetilde M^{S_ p}_{V-M_R}(\ell_i)
\otimes_m \widetilde M^{S_ p}_{V-M_L}(\ell_j)) \\
& \hspace{6cm} \oplus (\widetilde M^{(T_p)-S_ p}_{V-M_R}(\ell_i)
\otimes_e \widetilde M^{(S_p)-T_ p}_{V-M_L}(\ell_j))
\end{aligned}$

are the interaction fields between the $i$-th right semilepton and the $j$-th left semilepton: they decompose according to:
\Bean
\item {\bbf $(\widetilde M^{T_p-S_ p}_{V-M_R}(\ell_i)
\otimes_D \widetilde M^{T_p-S_ p}_{V-M_L}(\ell_j))$ which is a ``mixed'' diagonal space-time interaction field of gravitational nature\/} as developed in \cite{Pie2} and as it results from proposition 2.15 \cite{Ein2}.

\item {\bbf $(\widetilde M^{S_ p}_{V-M_R}(\ell_i)
\otimes_m \widetilde M^{S_ p}_{V-M_L}(\ell_j))$ which is a ``mixed'' magnetic ``space'' interaction field.}

\item {\bbf $(\widetilde M^{(T_p)-S_ p}_{V-M_R}(\ell_i)
\otimes_e \widetilde M^{(S_p)-T_ p}_{V-M_L}(\ell_j))$ which is a ``mixed'' electric ``time-space'' or ``space-time'' interaction field.}
\Ee
\Ei
\vskip 11pt

\subsection{Interacting (bisemi)photons}

{\bbf The (operator-valued) string fields at the ``~$ST$~'', ``~$MG$~'' and ``~$M$~'' levels of a set of ``~$K$~'' interacting bisemiphotons\/} are given, with reference to section 2.11, by:
\begin{multline*}
\phi ^{S_p}_{V-M_R}(p_K)\otimes
\phi ^{S_p}_{V-M_L}(p_K)\\
= \bigoplus^K_{i=1} (\phi ^{S_p}_{V-M_R}(p_i(\gamma _i))\otimes
\phi ^{S_p}_{V-M_L}(p_i(\gamma _i)))
\bigoplus^K_{i\neq j=1} (\phi ^{S_p}_{V-M_R}(p_i(\gamma _i))\otimes
\phi ^{S_p}_{V-M_L}(p_j(\gamma _j)))
\end{multline*}
where:
\Bi
\item $p_i(\gamma _i)$ denotes the $i$-th semiphoton at $\gamma _i$ quanta, $\gamma _i\in\NN$~.

\item $\phi ^{S_p}_{V-M_R}(\pt)$ \resp{$\phi ^{S_p}_{V-M_L}(\pt)$} are, in condensed form, the ``~$ST$~'', ``~$MG$~'' and ``~$M$~'' \rl strings (i.e., $\cit$-valued differentiable functions) of the considered semiphoton.

\item $\phi ^{S_p}_{V-M_R}(p_i(\gamma _i))\otimes \phi ^{S_p}_{V-M_R}(p_i(\gamma _i))$ are the ``~$ST$~'', ``~$MG$~'' and ``~$M$~'' bistrings at $\gamma _i$ biquanta of the $i$-th bisemiphoton decomposing into diagonal internal bistrings and into magnetic subfields at these levels responsible for the exchange of $\beta _i$ magnetic biquanta, $\beta _i\le \gamma _i$ according to section 2.11.

\item $\phi ^{S_p}_{V-M_R}(p_i(\gamma _i))\otimes
\phi ^{S_p}_{V-M_L}(p_j(\gamma _j))=
(\phi ^{S_p}_{V-M_R}(p_i(\gamma _i))\otimes_D
\phi ^{S_p}_{V-M_L}(p_j(\gamma _j)))\oplus\linebreak
(\phi ^{S_p}_{V-M_R}(p_i(\gamma _i)) \otimes_m
\phi ^{S_p}_{V-M_L}(p_j(\gamma _j)))$ are the {\bbf interaction fields between the $i$-th right semiphoton at $\gamma _i$ quanta and the $j$-th left semiphoton at $\gamma _j$ quanta\/}: they decompose according to:
\Bean
\item {\bbf a ``mixed'' diagonal   gravitational space interaction subfield\/}\linebreak
$(\phi ^{S_p}_{V-M_R}(p_i(\gamma _i))\otimes_D
\phi ^{S_p}_{V-M_L}(p_j(\gamma _j)))$~,
\item {\bbf a ``mixed'' magnetic interaction subfield\/}
$(\phi ^{S_p}_{V-M_R}(p_i(\gamma _i))\otimes_m
\phi ^{S_p}_{V-M_L}(p_j(\gamma _j)))$~.
\Ee
\Ei
\vskip 11pt

\subsection{Interaction fields of interacting (bisemi)baryons}

{\bbf The (operator-valued) string fields at the ``~$ST$~'', ``~$MG$~'' and ``~$M$~'' levels of a set of ``~$I$~'' interacting bisemibaryons\/} are given by:
\begin{multline*}
(\widetilde M^{(Bar);T_p-S_p}_{V-M_R}(b_I)
\otimes \widetilde M^{(Bar);T_p-S_p}_{V-M_L}(b_I))\\
= \bigoplus^I_{i=1} (\widetilde M^{(Bar);T_p-S_p}_{V-M_R}(b_i)
\otimes \widetilde M^{(Bar);T_p-S_p}_{V-M_L}(b_i))
\bigoplus^I_{i\neq j=1} (\widetilde M^{(Bar);T_p-S_p}_{V-M_R}(b_i)
\otimes \widetilde M^{(Bar);T_p-S_p}_{V-M_L}(b_j))
\end{multline*}
where:
\Bi
\item $(\widetilde M^{(Bar);T_p-S_p}_{V-M_R}(b_i)
\otimes \widetilde M^{(Bar);T_p-S_p}_{V-M_L}(b_i))$ are the ``~$ST$~'', ``~$MG$~'' and ``~$M$~'' internal string fields of the $i$-th bisemibaryon as developed in proposition 2.15.

\item $(\widetilde M^{(Bar);T_p-S_p}_{V-M_R}(b_i)
\otimes \widetilde M^{(Bar);T_p-S_p}_{V-M_L}(b_j))=
(\widetilde M^{*(Bar);T_p}_{V-M_R}(b_i)
\otimes \widetilde M^{*(Bar);T_p}_{V-M_L}(b_j))$

$\quad 
\bigoplus^3_{\alpha ,\beta =1}
(\widetilde M^{q_\alpha ;T_p-S_p}_{V-M_R}(b_i)
\otimes \widetilde M^{q_\beta ;T_p-S_p}_{V-M_L}(b_j))
\bigoplus^3_{\alpha =1}
(\widetilde M^{*(Bar);T_p-S_p}_{V-M_R}(b_i)
\otimes \widetilde M^{q_\alpha ;T_p-S_p}_{V-M_L}(b_j))$

$\quad
\bigoplus^3_{\alpha =1}
(\widetilde M^{q_\alpha ;T_p-S_p}_{V-M_R}(b_i)
\otimes \widetilde M^{*(Bar);T_p-S_p}_{V-M_L}(b_j))$
are the {\bbf ``mixed" interaction fields between the $i$-th right semibaryon ``~$b_i$~'' and the $j$-th left semibaryon ``~$b_j$~''.

These interaction fields are:\/}
\Bean
\item {\bf a ``strong'' gravitational field}
$(\widetilde M^{*(Bar);T_p}_{V-M_R}(b_i)
\otimes \widetilde M^{*(Bar);T_p}_{V-M_L}(b_j))$
between ``~$b_i$~'' and ``~$b_j$~'' responsible for the exchange of ``strong'' gravitational biquanta.

\item {\bbf gravito-electro-magnetic subfields}
$(\widetilde M^{q_\alpha ;T_p-S_p}_{V-M_R}(b_i)
\otimes \widetilde M^{q_\beta ;T_p-S_p}_{V-M_L}(b_j))$
{\bbf between the $\alpha $-th semiquark of the $i$-th right semibaryon ``~$b_i$~'' and the $\beta $-th semiquark of the $j$-th left semibaryon ``~$b_j$~''} according to proposition 2.15.

\item {\bbf ``strong'' gravitational and electric subfields}
$(\widetilde M^{*(Bar);T_p}_{V-M_R}(b_i)
\otimes\linebreak \widetilde M^{q_\alpha ;T_p-S_p}_{V-M_L}(b_j))$
\resp{$(\widetilde M^{q_\beta ;T_p-S_p}_{V-M_R}(b_i)
\otimes \widetilde M^{*(Bar);T_p}_{V-M_L}(b_j))$}
between the right core-time structure of ``~$b_i$~'' and the $\alpha $-th left semiquark of ``~$b_j$~'' \resp{the $\beta $-th right semiquark of ``~$b_i$~'' and the left core-time structure of  ``~$b_j$~''}
in such a way that the direct sums
\[ [
(\widetilde M^{*(Bar);T_p}_{V-M_R}(b_i)
\otimes \widetilde M^{q_\alpha ;T_p}_{V-M_L}(b_j))
\oplus
(\widetilde M^{q_\beta ;T_p}_{V-M_R}(b_i)
\otimes \widetilde M^{*(Bar) ;T_p}_{V-M_L}(b_j))]\]
of the {\bbf mixed gravitational ``strong'' string subfields}

and the direct sums
\[ [
(\widetilde M^{*(Bar);T_p}_{V-M_R}(b_i)
\otimes_e \widetilde M^{q_\alpha ;S_p}_{V-M_L}(b_j))
\oplus
(\widetilde M^{q_\beta ;S_p}_{V-M_R}(b_i)
\otimes_e \widetilde M^{*(Bar) ;T_p}_{V-M_L}(b_j))]\]
of the {\bbf mixed electric ``strong'' string subfields}

are {\bbf responsible for the generation of mesons of quark composition\/} $\o{q_\beta }q_\alpha $ as developed in proposition 2.15.
\Ee
\Ei
\vskip 11pt

\subsection{Interaction fields between interacting bisemileptons and bisemibaryons}

{\bbf The (operator-valued) string fields at the ``~$ST$~'', ``~$MG$~'' and ``~$M$~'' levels of a set of ``~$J$~'' bisemileptons interacting with a set of ``~$I$~'' bisemibaryons\/} are given by:
\begin{multline*}
[(\widetilde M^{T_p-S_p}_{V-M_R}(\ell_J)
\oplus \widetilde M^{(Bar);T_p-S_p}_{V-M_R}(b_I))
\otimes (\widetilde M^{T_p-S_p}_{V-M_L}(\ell_J)
\oplus \widetilde M^{(Bar);T_p-S_p}_{V-M_L}(b_I))]\\[11pt]
\begin{aligned}
&= \L[
\bigoplus^J_{i=1} \widetilde M^{T_p-S_p}_{V-M_R}(\ell_i)
\bigoplus^I_{k=1} \widetilde M^{(Bar);T_p-S_p}_{V-M_R}(b_k)\R]
\otimes
\L[
\bigoplus^J_{j=1} \widetilde M^{T_p-S_p}_{V-M_L}(\ell_j)
\bigoplus^I_{\ell=1} \widetilde M^{(Bar);T_p-S_p}_{V-M_R}(b_\ell)\R]\\[11pt]
&= \bigoplus^J_{i,j=1} 
( \widetilde M^{T_p-S_p}_{V-M_R}(\ell_i)
\otimes \widetilde M^{T_p-S_p}_{V-M_L}(\ell_j))
\bigoplus^I_{k,\ell=1} (\widetilde M^{(Bar);T_p-S_p}_{V-M_R}(b_k)
\otimes\widetilde M^{(Bar);T_p-S_p}_{V-M_R}(b_\ell))\\[11pt]
&\quad \L[
\bigoplus^J_{i=1} \bigoplus^I_{\ell=1}
(\widetilde M^{T_p-S_p}_{V-M_R}(\ell_i)
\otimes \widetilde M^{(Bar);T_p-S_p}_{V-M_L}(b_\ell))
\bigoplus^I_{k=1} \bigoplus^J_{j=1}
(\widetilde M^{(Bar);T_p-S_p}_{V-M_R}(b_k)
\otimes \widetilde M^{T_p-S_p}_{V-M_L}(\ell_j))\R]
\end{aligned}
\end{multline*}
where:
\Bi
\item $( \widetilde M^{T_p-S_p}_{V-M_R}(\ell_i)
\otimes \widetilde M^{T_p-S_p}_{V-M_L}(\ell_j))$~, $\forall\ i,j$=, $1\le i,j\le J$~,

are the {\bbf internal string fields\/} at the ``~$ST$~'', ``~$MG$~'' and ``~$M$~'' levels {\bbf of the $i$-th bisemilepton if $i=j$ and the gravito-electro-magnetic fields of interaction\/}  between the $i$-th right semilepton and the $j$-th left semilepton {\bbf if $i\neq j$\/} (see section 3.4);

\item $ (\widetilde M^{(Bar);T_p-S_p}_{V-M_R}(b_k)
\otimes\widetilde M^{(Bar);T_p-S_p}_{V-M_L}(b_\ell))$~, $\forall\ k,\ell$~, $1\le k,\ell\le I$~,

are similarly the  ``~$ST$~'', ``~$MG$~'' and ``~$M$~'' {\bbf internal string fields of the  $k$-th bisemibaryon if $k=\ell$ and the mixed interaction fields\/}  between the $k$-th right semibaryon ``~$b_k$~'' and the $\ell$-th left semibaryon ``~$b_\ell$~''{\bbf if $k\neq \ell$\/} according to section 3.6;

\item $(\widetilde M^{T_p-S_p}_{V-M_R}(\ell_i)
\otimes \widetilde M^{(Bar);T_p-S_p}_{V-M_L}(b_\ell))$

\quad $= (\widetilde M^{T_p-S_p}_{V-M_R}(\ell_i)
\otimes 
(\widetilde M^{*(Bar);T_p}_{V-M_L}(b_\ell)
\bigoplus^3_{\alpha =1} \widetilde M^{q_\alpha ;T_p-S_p}_{V-M_L}(b_\ell))$

\quad $= (\widetilde M^{T_p-S_p}_{V-M_R}(\ell_i)
\otimes 
\widetilde M^{*(Bar);T_p}_{V-M_L}(b_\ell)
\bigoplus^3_{\alpha =1} 
\widetilde M^{T_p-S_p}_{V-M_R}(\ell_i)
\otimes \widetilde M^{q_\alpha ;T_p-S_p}_{V-M_L}(b_\ell))$

are the {\bbf interaction fields\/} at the ``~$ST$~'', ``~$MG$~'' and ``~$M$~'' levels {\bbf between the $i$-th right semilepton ``~$\ell_i$~'' and the $\ell$-th left semibaryon ``~$b_\ell$~''\/} in such a way that:
\Bean
\item $(\widetilde M^{T_p-S_p}_{V-M_R}(\ell_i)
\otimes 
\widetilde M^{*(Bar);T_p}_{V-M_L}(b_\ell))$

\quad $=(\widetilde M^{T_p}_{V-M_R}(\ell_i)
\otimes_D
\widetilde M^{*(Bar);T_p}_{V-M_L}(b_\ell))
\oplus
(\widetilde M^{S_p}_{V-M_R}(\ell_i)
\otimes_e
\widetilde M^{*(Bar);T_p}_{V-M_L}(b_\ell))$

decomposes {\bbf into a gravitational field 
$(\widetilde M^{T_p}_{V-M_R}(\ell_i)
\otimes_D
\widetilde M^{*(Bar);T_p}_{V-M_L}(b_\ell))$ and into an electric field
$(\widetilde M^{S_p}_{V-M_R}(\ell_i)
\otimes_e
\widetilde M^{*(Bar);T_p}_{V-M_L}(b_\ell))$~;}
\vskip 11pt

\item $(\widetilde M^{T_p-S_p}_{V-M_R}(\ell_i)
\otimes
\widetilde M^{q_\alpha ;T_p-S_p}_{V-M_L}(b_\ell))$
generates {\bbf a mixed gravito-electro-magnetic field of interaction\/} between the $i$-th right semilepton ``~$\ell_i$~'' and the $\alpha $-th left semiquark ``~$q_\alpha $~'' of the $\ell$-th left semibaryon ``~$b_\ell$~''.
\Ee
\Ei
\vskip 11pt

\subsection{Proposition}
{\em \Bena
\item A set of ``~$J$~'' bisemileptons interact between themselves by means of a gravito-electro-magnetic field.

\item A set of ``~$K$~'' bisemiphotons interact between themselves by means of a gravito-magnetic field.

\item A set of ``~$I$~'' bisemibaryons interact between themselves by means of:
\Be
\item a strong gravitational field between right and left core central structures of different bisemibaryons.

\item gravito-electro-magnetic fields between right and left semiquarks of different bisemibaryons

\item strong gravitational and electric fields between \rl core time structures and \lr semiquarks of different bisemibaryons.
\Ee
\Ee}
\vskip 11pt

\bpr These assertions result from the developments of sections 3.4, 3.5 and 3.6.\epr
\newpage

\subsection*{B) Implicit description of  interacting bisemiparticles by means of (bi)connections}

\subsection{Deformation of a bistring under an external potential field}

\Bi
\item The second way of handling the interactions between a set of ``~$J$~'' bisemiparticles consists in considering that one of these bisemiparticles is submitted to the global influence of the bilinear external field (operator $(A_R\times A_L)$~) of the subset of the $(J-1)$ remaining ``external'' bisemiparticles in such a way that every $(\mu ,m_\mu )$-th ``space'' bisection of the ``~$M$~'' level (as for the ``~$ST$~'' and ``~$MG$~'' levels) of this  considered bisemiparticle can join ``~$k$~'' external biquanta transforming it under {\bf the deformation\/} according to:
\begin{align*}
\Ds^{[\mu ]\to[\mu +k]_{S_p}}\RL: \quad
& \begin{aligned}[t]
&\phi ^{S_p}_{M_R}(M_{L_{\o v_{\mu ,m_\mu }}}) \otimes
\phi ^{S_p}_{M_L}(M_{L_{v_{\mu ,m_\mu }}}) \\
&\quad
\equiv (T^S_{R;M_\mu }\otimes T^S_{L;M_\mu })
(\phi ^{S}_{M_R}(M_{L_{\o v_{\mu ,m_\mu }}}) \otimes
\phi ^{S}_{M_L}(M_{L_{v_{\mu ,m_\mu }}}))
\end{aligned}\\
& \qquad \To \begin{aligned}[t]
&\phi ^{S_p}_{M_R}(M_{L_{\o v_{(\mu+k) ,m_{(\mu+k)} }}}) \otimes
\phi ^{S_p}_{M_L}(M_{L_{v_{(\mu +k),m_{(\mu+k)} }}}) \\
&\quad
\equiv [(T^S_{R;M_\mu }+\vec A_R(r))\otimes (T^S_{L;M_\mu }+\vec  A_L(r))]\\
&\qquad
[(\phi ^{S}_{M_R}(M_{L_{\o v_{(\mu+k),m_{(\mu+k)} }}}) \otimes
\phi ^{S}_{M_L}(M_{L_{v_{(\mu+k) ,m_{(\mu+k)} }}}))]\;, \\
& \hspace{7cm} 0\le k\le \infty \;,
\end{aligned}\\
\end{align*}

where:
\Bi
\item $\Ds^{[\mu ]\to[\mu +k]_{S_p}}\RL$ is a deformation similar to these introduced in section 2.11.

\item $T^S_{L;M_\mu }$ \resp{$(T^S_{R;M_\mu }$} is the \lr linear momentum operator at the ``mass'' level given in proposition 2.9.

\item $\vec A_L(r)$ \resp{$\vec A_R(r)$} is the \lr external field potential operator acting on the \lr section $\phi ^{S}_{M_L}(M_{L_{v_{k ,m_k }}})$ \resp{$\phi ^{S}_{M_R}(M_{L_{\o v_{k ,m_k }}})$} at $k$ quanta.
\Ei
\vskip 11pt

\item The ``added'' ``~$k$~'' biquanta proceed from the external field at $(J-1)$ bisemiparticles and are not necessarily connected to the $\mu $ biquanta of the (operator-valued) bistring $\phi ^{S_p}_{M_R}(M_{L_{\o v_{\mu ,m_\mu }}}) \otimes
\phi  ^{S_p}_{M_L}(M_{L_{v_{\mu ,m_\mu }}})$~.

Indeed, we have that:
\[ \phi ^{S}_{M_L}(M_{L_{v_{(\mu+k) ,m_{(\mu+k)} }}})
\simeq 
\phi ^{S_p}_{M_L}(M_{L_{v_{\mu ,m_\mu }}}) \oplus
\phi ^{S_p}_{M_L}(M_{L_{v_{k ,m_k }}})\]
(idem for the right case).

Remark that this transformation corresponds in quantum (field) theory to the invariance of the wave function under a phase factor.
\vskip 11pt

\item If, instead of envisaging an (operator-valued) bisection (or bistring) of ``space'', we take into account an (operator-valued) bisection of ``space-time'' (see section 2.3), then the external field operator to be considered is the generic biconnection $(A_R(t,r)\otimes A_L(t,r))$ where $A_L(t,r)$ \resp{$A_R(t,r)$} is a \lr connection acting on a \lr time space string and being a \lr distribution at each \lr point of it.

$A_L(t,r)$ \resp{$A_R(t,r)$} is a four-vectorial distribution:
\begin{align*}
A_L(t,r) &= \{A^t_L,A^x_L,A^y_L,A^z_L\}\\
\rresp{A_R(t,r) &= \{A^t_R,A^x_R,A^y_R,A^z_R\}}
\end{align*}
of which $\vec A_L(r)$ \resp{$\vec A_R(r)$} may be given by \cite{A-L}, \cite{B-J}.
\begin{align*}
\vec A_L(r) &= \int d^3k_L\ A_L(\vec k,\vec s)\ e^{i\vec k\vec r}\ \varepsilon (k_L,\lambda )\\
\rresp{\vec A_R(r) &= \int d^3k_R\ A_R(\vec k,\vec s)\ e^{-i\vec k\vec r}\ \varepsilon (k_R,\lambda )}
\end{align*}
where $\varepsilon (k_L,\lambda )$ \resp{$\varepsilon (k_R,\lambda )$} is the polarization 
unit vector depending on the integer $\lambda =1,2$ referring
to the two transverse polarization modes of the semiphotons.

The intergral bears on the normal modes ``~$k$~'' referring to the number of ``~$k$~'' external quanta joining the section $\phi ^{S}_{M_L}(M_{L_{v_{\mu ,m_\mu }}})$ \resp{$\phi ^{S}_{M_R}(M_{L_{\o v_{\mu ,m_\mu }}})$} at the mass level of the considered bisemifermion under the action of the external vector potential $\vec A_L(r)$ \resp{$\vec A_R(r)$}.
\vskip 11pt

\item Let $\MM_R$ \resp{$\MM_L$} denote the four-vectorial mass operator $T^{T-S}_{R;M_\mu }$ \resp{$T^{T-S}_{L;M_\mu }$} introduced in proposition 2.9 and let $A_R$ \resp{$A_L$} be the four-vectorial distribution $A_L(t,r)$ \resp{$A_R(t,r)$}.
\Ei
\vskip 11pt

Then, we can state the following proposition:
\vskip 11pt

\subsection{Proposition}

{\em \Bena
\item The mass bioperator $(\MM_R\otimes \MM_L)$~, acting on every mass bisection (or bistring) of an elementary bisemifermion and endowed with the infinitesimal biconnection\linebreak $(eA_R\otimes eA_L)$~, noted $(A_R\otimes A_L)$~, will develop according to:
\[ (\MM_R+A_R)\otimes (\MM_L+A_L)=(\MM_R\otimes \MM_L)+(A_R\otimes A_L)
+ [(\MM_R\otimes A_L)+(A_R\otimes \MM_L)]\]
where $[(\MM_R\otimes A_L)+(A_R\otimes \MM_L)]$ is the {\bfseries gravito-electro-magnetic interaction field operator\/} of which tensorial form is:
\[ \MM A_{mn}=\MM_mA_n+A_m\MM_n\;, \qquad m,n=t,x,y,z\;. \]
The explicit form of this $GEM$ tensor is:
\[\MM A_{mn}=\begin{bmatrix}
G_t & E^-_x & E^-_y & E^-_z\\
 E^+_x & G_x & B^-_z & B^+_y\\
 E^+_y &  B^+_z & G_y & B^-_x\\
 E^+_z &  B^-_y &   B^+_x & G_z\end{bmatrix}\]
 where:
 \Be
 \item $\vec E^{\pm}=\{E^\pm_x,E^\pm_y,E^\pm_z\}$ is a $3D$-positively \resp{negatively} charged electric field operator.
 
 \item $\vec B^{\pm}=\{B^\pm_x,B^\pm_y,B^\pm_z\}$ is a $3D$-positive \resp{negative}  magnetic field operator.
 
\item $G=[G_x,G_y,G_z]$ is a $3D$-gravitational field diagonal tensor and $G_t$ is the  ``time'' scalar gravitational field.
\Ee
\vskip 11pt

\item The symmetric $GEM$ tensor $\MM A_{mn}$ is transformed into the antisymmetric tensor $F_{mn}$ of electromagnetism if $\MM A_{mn}$ is submitted to the bijective antisymmetric map\linebreak $C: \MM A_{mn}\to F_{mn}$ transforming the right components $A_m$ into their corresponding left components and the left components $\MM_n$ into their corresponding right components.
\Ee
}
\vskip 11pt

\bpr
\Bi
\item The gravito-electro-magnetic interaction field operator $GEM$ as well as its one-to-one correspondence with the antisymmetric tensor $F_{mn}$ of electromagnetism was introduced and proved in \cite{Pie2}.

\item However, let us recall that the off-diagonal electric components $E^-_i$ of the $GEM$ tensor $\MM A_{mn}$ are given by:
\[ E^-_i \simeq m_0\ A_i+ A_t\ p_i\simeq i\ \hbar\ \F \partial{\partial t}\ A_i-A_t\cdot i\ \F\hbar c\ \F\partial{\partial i}\;, \quad i=x,y,z\;, \]
if $m_0$ and $p_i$ are the condensed notations respectively for the $\mu $-th proper mass operator $s_{0_{R_\mu }}\ \dfrac\partial{\partial t_{0_\mu }}$ and the $\mu $-th linear momentum operator components $s_{i_{L_\mu }}\ \dfrac\partial{\partial i_\mu }$ of the considered elementary bisemifermion (see proposition 2.9).

\item The antisymmetric one-to-one correspondence $C:\MM A_{mn}\to F_{mn}$ transforms $\{E^-_i\}_i$ into:
\[ C: \quad E^-_i \To -E_i\;, \qquad i=x,y,z\;, \quad E_i\in F_{m,n}\;,\]
in such a way that:
\[ -E_i = i\ \L(\F{\partial A_i}{\partial t}-\F{\partial A_t}{\partial i}\R)\]
where the $c=\hbar=1$ unit system is considered as well as the map: $C:s_{m_{L,R_\mu }}\to 1$ (~$L,R$ meaning ``left'' or ``right'').

\item Let us remark that the novelty of the symmetric tensor $\MM A_{mn}$ with respect to the antisymmetric tensor $F_{mn}$ of electromagnetism is {\bf the existence of diagonal components}
\begin{align*} 
G_t&\simeq m_0\ A_t + A_t\ m_0 \\
 \text{and} \quad G_i&\simeq p_i\ A_i+A_i\ p_i\end{align*}
{\bbf which belong respectively to a scalar ``time'' gravitational field and a ``~$3D$-space'' gravitational field diagonal tensor\/}.

Under the antisymmetric map $C$~, these gravitational components become null.\epr
\Ei
\vskip 11pt

\subsection{Corollary}

{\em The ``~$GEM$~'' gravito-electro-magnetic tensor $\MM A_{mn}$ is reduced to the 
``~$GM$~'' gravito-magnetic tensor $\MM A^p_{ij}$~, $i,j=x,y,z$~, in the case of bisemiphotons, i.e. when a bisemiphoton interacts with an external field.}
\vskip 11pt

\bpr As the bisemiphotons interact by means of a gravito-magnetic field according to proposition 3.8, the $\MM A_{mn}$ tensor reduces to the gravito-magnetic tensor
\[ \MM A^p_{ij}=\begin{bmatrix}
G_x & B^-_z & B^+_y \\
 B^+_z & G_y &B^-_x \\
 B^-_y & B^+_x & G_z\end{bmatrix}\]
 which is transformed under the antisymmetric map $C$ into the antisymmetric tensor:
 \be F^p_{ij}= \begin{bmatrix}
 0 & -B_z & +B_y\\
 +B_z & 0 & -B_x\\
 -B_y & +B_x & 0\end{bmatrix}\;.\tag*{\eop}\ee
 \vskip 11pt

\subsection{Proposition}

{\em
\Bena
\item {\bfseries The condition of {\boldmath  $4D$-null divergence $\partial^n\MM A_{mn}=0$}~, i.e.}
\[ (1\otimes \delta _L)[(\MM_R\otimes A_L)+(A_R\otimes \MM_R)]=0\;, \]
{\bfseries applied to the {\boldmath $GEM$} tensor {\boldmath $\MM A_{mn}$} leads to the set of {\boldmath $GEM$} differential equations which are extended equations of Maxwell:}
\[\begin{cases}
\vec \nabla \centerdot \vec E &= \dfrac{\partial G_t}{\partial t}\;, \\[11pt]
\vec \nabla \times \vec B &=\vec \nabla \ G+\dfrac{d\vec E}{dt}\;, \end{cases}\]
analog to the second set of Maxwell equations $\partial^n F_{mn}=4\pi j_m$ or
\[\begin{cases}
\vec \nabla \centerdot \vec E &= \rho \;, \\[11pt]
\vec \nabla \times \vec B &=\vec j+\dfrac{d\vec E}{dt}\;, \end{cases}\]
where $j_m$ are the $t,x,y,z$ components of the $4D$-current $\{\rho ,\vec j\}$~.
\vskip 11pt

\item {\bfseries The gravito-electro-magnetic differential equations:
\[\begin{cases}
\vec \nabla \centerdot \vec E &= \dfrac{\partial G_t}{\partial t}\;, \\[11pt]
\vec \nabla \times \vec B &=\vec \nabla \ G+\dfrac{d\vec E}{dt}\;, \end{cases}\]
describe} no more, as in the Maxwell equations,
{\bfseries the flux of {\boldmath  $\vec E$ through a closed surface $(\vec \nabla\pt\vec E)$ or the circulation of $\vec B$ around a loop, respectively,}}
\Bi
\item by means of the charge density inside and the current through the loop,
\item but, {\bfseries in function of the variation in time of the time gravitational {\boldmath  $G_t$ and of the flux of the space gravitational field $G$ through the loop}.}
\Ei
{\bfseries This allows to unify the electromagnetism of Maxwell with the gravitation as hoped by A. Einstein} \cite{Ein1}.\Ee
}
\vskip 11pt

\bpr
\Bi
\item When the second set of the Maxwell equations describes the magneto-electrodynamics in function of the electric charge density and the electric current, {\bbf $GEM$ differential equations give an explanation of how the gravito-magneto-electrodynamics works: this gives} theoretically {\bbf the possibility of generating a $1D$ and $3D$ gravitational field} respectively from an electric field and from an electromagnetic field (or vice-versa).

\item Remark that a gravitational field is only generated if the ``added'' ``~$k$~'' biquanta proceeding from the external fields are not connected to the considered bisemiparticle as resulting from the developments of section 3.9.

\item Finally, the conditions of $4D$-null divergence $\partial^n\MM A_{mn}=0$ of the $GEM$ tensor $\MM A_{mn}$ leads to the conditions $(\partial_L,A_L)=(\partial_L,\MM_L)=0$~, where $\partial _L$ is a $4D$-(left-) divergence and $(\pt,\pt)$ is a scalar product.

Now, $(\partial_L,A_L)=0$ corresponds to the radiation gauge or to the Lorentz condition of electro-magnetism while $(\partial_L,\MM_L)=0$ is a condition of conservation of the left mass of the reference left semiparticle or is a condition of uniform (non accelerated) motion.\epr
\Ei

\section{The Feynman paths in AQT}

Taking into account the new way of considering the (bilinear) interactions between (bisemi)\-particles \cite{Pie1} as developed in chapters 2 and 3, the Feynman paths will be adjusted and reinterpreted in this chapter at the light of the developments of string theory and quantum field theory.
\vskip 11pt

\subsection{Three kinds of gravito-electro-magnetic bilinear interactions}

\def\ST{``~$ST$~''\xspace}
\def\MG{``~$MG$~''\xspace}
\def\M{``~$M$~''\xspace}

Let us outline that {\bbf three kinds of bilinear interactions at the \ST, \MG and \M levels\/} have been envisaged:
\Bena
\item {\bbf Internal explicit of (gravito-)electro-magnetic type\/} allowing to join the left and right semiparticles of a given bisemiparticle. 

More concretely:
\Bi
\item  a bisemifermion is tied by means of an internal electro-magnetic field (see proposition 2.8).
\item a bisemiphoton holds together by a magnetic subfield (see proposition 2.11).
\item a bisemibaryon is stuck by means of the internal electro-magnetic subfields of the bisemiquarks, the gravito-electro-magnetic subfields of interaction between the left and right semiquarks of the different bisemiquarks and, finally, by the mixed ``strong'' gravitational and electric subfields of interaction between the core central time structures of the left and right semibaryons and, respectively, the right and left semiquarks (see proposition 2.15).
\Ei

\item {\bbf External explicit of gravito-electro-magnetic type\/} allowing to describe the bilinear interactions between a set of ``~$J$~'' interacting bisemiparticles by means of gravito-electro-magnetic subfields between left and right semiparticles belonging to different bisemiparticles (see proposition 3.8).

\item {\bbf External implicit by means of infinitesimal biconnections\/} allowing the introduction of a gravito-electro-magnetic interaction field tensor in one-to-one correspondence with the antisymmetric tensor of Maxwell.
\Ee

These three kinds of bilinear interactions are of gravito-electro-magnetic type and are produced by the exchange of gravitational, electric and magnetic  biquanta, the exchanged bisemibosons generating gravito-electro-magnetic forces between right and left semiparticles.

{\bbf The exchange of electro-magnetic bisemibosons\/}, transferring momenta, {\bbf is responsible for the Coulomb force\/} and these electro-magnetic bisemibosons are the virtual photons of electrodynamics.

What is new in this algebraic quantum theory is that the exchange of {\bbf gravitational biquanta, responsible for the gravitational force\/}, is also considered on an equal footing as the exchange of electro-magnetic biquanta.
\vskip 11pt

\subsection{The Feynman paths and the perturbative series in QFT}

In QFT, the Feynman paths describe, among other things, the exchange of virtual photons by studying {\bf functional integrals\/}
\[ \int_{{\rm Map}(\Sigma ,M)} F(\phi )\ e^{-\F i\hbar S(\phi )}\ D\phi \]
where:
\Bi
\item $\phi $ is a map from a space-time manifold ``~$\Sigma $'' into a target space ``~$M$~'', which may be $\rit$~: so, ${\rm Map}(\Sigma ,M)$ may be considered as a Hilbert space.
\item $D\phi $ is a product of local measures on ``~$M$~''.
\item $F(\phi )$ is a functional of $\phi $ and $S(\phi )$ the corresponding action.
\Ei

Such a functional integral may be given by the correlation function depending on a sequence $\{x_i\}^n_{i=1}$ of points in $\Sigma $ \cite{Gaw2}, \cite{Ben}, \cite{Bou}:
\[ \langle \phi (x_1)\ \dots\ \phi (x_n)\rangle
= \F{\int_{{\rm Map}(\Sigma ,M)} \phi (x_1)\ \dots\ \phi (x_n)\ e^{-S(\phi )}\ D\phi}
{\int_{{\rm Map}(\Sigma ,M)} \ e^{-S(\phi )}\ D\phi}\;.\]

These functional integrals appear when the $S$-matrix is calculated by {\bbf perturbative techniques\/} introduced by Feynman, Schwinger and Tomonaga.

A unified approach of these perturbative techniques was proposed  by Dyson as {\bbf a time dependent perturbation theory\/} which can be succintly introduced as follows \cite{Dys}:

Consider the differential equation
\[ i\ \hbar\ c\ \F{\partial\psi (\sigma )}{\partial\sigma }
= V(\sigma )\ \psi (\sigma )\]
where:
\Bi
\item $V(\sigma )$ is the time dependent interaction potential of interacting fields;
\item $\psi (\sigma )=U(\sigma ,\sigma _0)\ \psi _0$ is defined over a family $\sigma _0,\sigma _1,\dots$ of surfaces.
\Ei

{\bbf The operator $U(\sigma ,\sigma _0)$\/}, satisfying the differential equation
\[ i\ \hbar\ c\ \F{\partial U(\sigma ,\sigma _0)}{\partial\sigma }
= V(\sigma )\ U(\sigma,\sigma _0 )\;,\]
can be put in the form
\[
U(\sigma ,\sigma _0)
= \L( 1-\F{i\ \hbar}c\ \int^{\sigma_1}_{\sigma _0} V(t)\ dt\R)
\times  \L( 1-\F{i\ \hbar}c\ \int^{\sigma_2}_{\sigma _1} V(t)\ dt\R)
\times \dots\]
which, {\bbf expanded in ascending powers of $V$\/} (and of $\hbar$~) gives:
\begin{align*}
U(\sigma,\sigma _0 ) =
&1-\F{i\ \hbar}c\ \int^{\sigma}_{\sigma _0} dt_1\ V(t_1)\\
&\quad \L(-\F{i\ \hbar}c\R)^2\ \int^{\sigma}_{\sigma _0} dt_1\ \int^{\sigma(t_1)}_{\sigma _0} dt_2\  V(t_1)\ V(t_2)\\
&\quad \L(-i\F{\hbar\ c}c\R)^3\ \int^{\sigma}_{\sigma _0} dt_1\ \int^{\sigma(t_1)}_{\sigma _0} dt_2\  \int^{\sigma(t_2)}_{\sigma _0}\ dt_3V(t_1)\ V(t_2)\ V(t_3)\\
& \quad + \dots\;.
\end{align*}
The $V(t_i)$~, $1\le i\le n\le \infty $~, thus are scattering potentials due to the presence of ``~$n$~'' external particles or ``~$n$~'' external sets of particles and the form
\[
U(\sigma ,\sigma _0)
= \L( 1-\F{i\ \hbar}c\ \int^{\sigma_1}_{\sigma _0} V(t)\ dt\R)
\times  \L( 1-\F{i\ \hbar \ c}c\ \int^{\sigma_2}_{\sigma _1} V(t)\ dt\R)
\times \dots\]
of $U(\sigma ,\sigma _0)$ is directly related to the fact that the free field is expressed as a(n) (anti)-symmetric product of the individual fields.
\vskip 11pt

We can thus say that {\bbf the perturbative series in QFT arise as a consequence of the development of the free fields in products of the individual fields\/}.
\vskip 11pt

\subsection{The interactions in AQT are not worked out perturbatively}

When QFT works linearly at a fundamental level with amplitudes \cite{Fey1}, the algebraic quantum theory (AQT),  being a bilinear theory, works bilinearly with intensities by means of representations of bilinear (algebraic) semigroups \cite{Pie2}, \cite{Pie3}.

So, in AQT, a set of \mfg{J} non interacting fields (i.e. free fields) at the ``space'' ``mass'' level is given according to section 3.1 and proposition 3.2 by:
\begin{align*}
\FRepsp(\GL_{2J}(L_{\o V}\times L_V))
&= \L(\Boxplus^J_{i=1}(\FRepsp T^t_2(L_{\o v_i}))\R)
\otimes \L(\Boxplus^J_{i=1}(\FRepsp T_2(L_{v_i}))\R)\\
&= \bigoplus^J_{i=1}\L(\widetilde M^{S_{p}}_{M_R}(i) \otimes \widetilde M^{S_{p }}_{M_L}(i)\R)
\end{align*}
where $\FRepsp(\centerdot)$ is a functional representation space.

It thus appears clearly that the mass free field is given by the sum, over the \mfg{J} considered bisemiparticles, of the tensor products of the right semifields $M^{S_{p }}_{M_R}(i)$ by the left semifields $M^{S_{p}}_{M_L}(i)$~.

In other respects, the ``space'' ``mass'' interaction fields are given by:
\[
\Boxplus^J_{i\neq j=1}\L(\FRepsp T^t_{2_i}(L_{\o v_i}))
\otimes \FRepsp T_{2_j}(L_{v_j}))\R)
= \bigoplus^J_{i\neq j=1}\L(\widetilde M^{S_{p}}_{M_R}(i) \otimes \widetilde M^{S_{p }}_{M_L}(j)\R)\;.
\]

As a consequence, {\bbf the interaction between fields cannot be expressed in AQT at a fundamental level by perturbative series on the contrary of what is done in QFT.}

All that is contained in the following proposition.
\vskip 11pt

\subsection{Proposition}

{\em The interactions between fields are not realized in AQT at a fundamental level by perturbative series.}
\vskip 11pt

\bpr Indeed, AQT, being a bilinear theory, works at a fundamental level with intensities.  This results from the fact that the free fields are given in AQT by the reducible orthogonal functional representation space $\FRepsp \GL_{2J}(L_{\o V}\times L_V)$ of the bilinear algebraic semigroup $\GL_{2J}(L_{\o V}\times L_V)$ while the interaction fields are expressed as the reducible non orthogonal functional representation space of the same bilinear algebraic semigroup.

Consequently, the interactions between fields are not realized in a perturbative way as it was justified in section 4.3.\epr
\vskip 11pt

\subsection{Transitions amplitudes of the Feynman paths}

\Bi
\item On the other hand, each \mfg{ST}, \mfg{MG} or \mfg{M} (operator-valued) field of space or time in AQT is given by (the sum of) the set of the products, right by left, of its sections, which are strings characterized by their increasing ranks $[L_{v_{\mu ,m_\mu }}:K]\simeq \mu \centerdot N$ or ``Fourier'' modes \mfg{\mu }.

Each bisection of normal (or ``Fourier'') mode \mfg{\mu } is then  a bistring behaving like a harmonic oscillator.

{\bbf We are thus led to study the interactions at the level of the strings in the philosophy of the Feynman paths in a non perturbative way.}

\item If we refer now to the {\bbf transition amplitude\/} obtained in QFT for a single point particle traveling from a point \mfg{x} to a point \mfg{x'} of a Riemannian manifold $M$~, we know that it is expressible {\bbf in terms of the sum of all paths\/} joining $x$ to $x'$ \cite{D'Ho}, \cite{Wein}:
\[ S(x';x) = \sum_{\substack{{\rm paths}\\ x\to x'}} e^{-S[{\rm path}]}\;.\]

\item This transition amplitude $S({\mathbf{x}'};{\mathbf{x}})$ can be worked out by considering {\bbf the evolution in time of the (non relativistic) wave function $\psi (x,t)$ of the point particle\/} as given by:
\[ \psi ({\mathbf{x}'},t') = i\int d^3x\ G({\mathbf{x}'},t';{\mathbf{x}},t)\ \psi ({\mathbf{x}},t)\;, \qquad {\mathbf{x}},{\mathbf{x}'}\in\rit^3\;, \quad t'>t\;, \]
where:
\Bi
\item $G({\mathbf{x}'},t';{\mathbf{x}},t)$ is the Green's propagator;
\item $\psi ({\mathbf{x}'},t')$ is in fact a wave \cite{B-J} depending on the interval $[({\mathbf{x}'}-{\mathbf{x}}),(t'-t)]$~.
\Ei
\Ei
\vskip 11pt

\subsection{Feynman path intervals for bistrings}

\Bi
\item In the philosophy of AQT, the left and right strings are one-dimensional waves, homotopic to circles $S^1$ and characterized by two senses of rotation according to section~2.3.

Referring to the evolution in time of the wave function at a point particle, we can easily state that:
\Bena
\item {\bbf The one-dimensional string waves\/} of AQT {\bbf have an evolution in time which depends:}
\Be
\item {\bbf either on the intervals\/} $[(x'-x),(t'-t)]$, where $x,x'\in\rit^1$~,
\item or on a transition from a normal mode $\mu $ (at $\mu $ quanta) at \mfg{x} to a normal mode $\nu $ (at $\nu $ quanta) at \mfg{x'}~.
\Ee

\item As the string waves rotate, the path intervals $[x'(t)-x(t)]$ of these correspond in fact to rotations of some angles of these strings.

So, the {\bbf Feynman path intervals in AQT are fundamentally arcs of circles measured by angles of rotation\/} of the considered string waves which are rotated by Green's propagators.
\Ee

\item But, as the string fields of AQT are composed of bistrings of which left and right strings do not have necessarily the same rotational velocity, exchanges of magnetic quanta occur between them provoking a global movement of translation of the bistrings as developed in \cite{Pie2}.

So, {\bbf a left and a right string of a given bistring give rise theoretically to two adjacent world sheets\/} covering possibly each other partially.  But, the world sheets \cite{H-S-V}, \cite{G-S}, \cite{Moo}, \cite{Poly} of string theory \cite{Gid}, \cite{Pol1}, \cite{Pol2}, \cite{Joh}, \cite{DelToWit} (as well as the world-lines of point particles) have a continuous character, which is not necessarily the case in the context of AQT, since {\bbf the associated left and right strings exchange ``discontinuously'' (magnetic) quanta\/}, allowing them to interact.

As a consequence, {\bbf the normal ``Fourier'' modes of the left and right strings\/} of a given bistring {\bbf change during their evolution in time\/}, and, thus, it does not seem realistic to envisage the world sheet of a string at a given normal mode \mfg{\mu }~.
\Ei
\vskip 11pt

\subsection{Proposition}

{\em The only basic diagrams of interaction in AQT are those involving exchanges of gravitational, electric and magnetic biquanta between left and right strings (or sets of strings): they correspond to the first order diagrams of Feynman.}
\vskip 11pt

\bpr 
\Bi
\item As the internal structure of the elementary bisemiparticles is composed of ``space'' and ``time'' fields at the \mfg{ST}, \mfg{MG} and \mfg{M} levels and as these fields are composed of bistrings, the basic diagrams of interaction in AQT are those involving left and right strings (or sets of strings).

\item As AQT is fundamentally a non perturbative theory of interaction, the only basic diagrams are those involving left and right strings (or sets of strings) exchanging gravitational, electric and magnetic biquanta as it was developed in chapter 2 and 3.

The corresponding diagrams of Feynman are those of first order of quantum electrodynamics implying the exchange of virtual photons \cite{Fey2}.

\item These diagrams describe the (inverse) rotations of left and right strings in such a way that the covered arcs of circles imply corresponding translations of the bistrings as recalled in section 4.6.\epr
\Ei
\vskip 11pt

\subsection{The basic diagrams of AQT}

These can be classified in three categories corresponding to the three kinds of bilinear interactions described in section 4.1.

They are:
\Bena
\item {\bf Internal explicit diagrams}

\Be
\item {\bf of magnetic type\/}  involving pairs of left and right strings belonging to the internal fields \mfg{ST}, \mfg{MG} and \mfg{M} of ``space'' (or ``time'') of the elementary bisemifermions and {\bbf describing the exchanges of internal magnetic biquanta ``inside'' these bistrings.}

\item {\bf of electric type\/} involving pairs of space (resp. time) left strings and time (resp. space) right strings belonging respectively to the internal space (resp. time) left semifields and to the internal time (resp. space) right semifields \mfg{ST}, \mfg{MG} and \mfg{M} of the elementary bisemifermions and {\bf describing the exchange of internal electric biquanta,} responsible for the electric charge at these levels \mfg{ST}, \mfg{MG} and \mfg{M}.

\item {\bf of gravitational type\/}
\Be
\item involving pairs of left and right strings belonging respectively to left and right semifields \mfg{ST}, \mfg{MG} and \mfg{M} {\bbf of left and right semiquarks of different bisemiquarks} of bisemibaryons and {\bf describing the exchanges of gravitational biquanta} between these pairs of strings.

\item involving pairs of left and right strings belonging respectively to the left and right ``time'' semifields \mfg{ST}, \mfg{MG} and \mfg{M} of semiquarks and of core central ``time'' semistructures of semibaryons and {\bf describing the exchanges of ``strong'' gravitational biquanta} inside bisemibaryons.
\Ee
\Ee

\item {\bf External explicit diagrams of magnetic, electric and gravitational types\/} as in 1) except that the envisaged pairs of left and right strings are such that the \lr and \rl strings belong to the semifields \mfg{ST}, \mfg{MG} and \mfg{M} of two different (elementary) bisemiparticles.

\item {\bf External implicit diagrams of magnetic, electric and gravitational types\/} as in 2) by means of infinitesimal biconnections implying that the left or right strings of the considered pairs of strings are now acted by the left or right external vector potentials $A_L$ or $A_R$ and result from deformations as developed in section 3.9.
\Ee

\Bi
\item So, the internal explicit diagrams refer to {\bbf the exchanges of gravito-electro-magnetic biquanta inside elementary bisemiparticles\/}.

\item  While the external explicit and implicit diagrams describe {\bbf the exchanges of gravito-electro-magnetic biquanta between different bisemiparticles allowing them to interact.}
\Ei
\vskip 11pt

\subsection{Bistrings in the diagrams of AQT}

{\bbf The basic diagrams in AQT involve products of right strings by left strings\/} of the types:
\[ \Phi ^S_{R_\mu }(x_R)
\otimes \Phi ^S_{L_\mu }(x_L)\]
where the \lr string $\Phi ^S_{L_\mu }(x_L)$ \resp{$\Phi ^S_{R_\mu }(x_R)$}, being the condensed notation of $\Phi ^S_{L}(M_{L_{v_{\mu ,m_\mu }}})$
\resp{$\Phi ^S_{R}(M_{L_{\o v_{\mu ,m_\mu }}})$} envisaged in section 2.4, will be written {\bbf in the automorphic representation (see \cite{Pie4}) of the bilinear algebraic semigroup\/} $\GL_2(L_{\o v}\times L_v)$ according to:
\[ \phi ^S_{L_\mu }(x_L)=c_\mu \ e^{2\pi i\mu x_L}
\rresp{\phi ^S_{R_\mu }(x_R)=c_\mu \ e^{-2\pi i\mu x_R}}\;.\]

{\bbf The amplitude $c_\mu $ \resp{$c^*_\mu $} of the string\/}
$\phi ^S_{L_\mu }(x_L)$ \resp{$\phi ^S_{R_\mu }(x_R)$} is (approximately) the radius of the circle $c_\mu \ e^{2\pi i\mu x_L}$ \resp{$c^*_\mu \ e^{-2\pi i\mu x_R}$} at $\mu $ quanta rotating counterclockwise (resp. clockwise).

The amplitudes $c_\mu $ and $c^*_\mu $ thus do not have in AQT the interpretation of creation an annihilation operators that they have in QFT.

Indeed, {\bbf the creation and annihilation operators\/} are given in AQT by deformations and inverse deformations of Galois type \cite{Maz} of strings based respectively on Galois automorphisms or antiautomorphisms as it was developed in \cite{Pie2} and in \cite{Pie4} (see quantization rules).
\vskip 11pt

\subsection{Proposition}

{\em The \lr {\bfseries Green's propagator of a \lr string is given by the one-parameter group of diffeomorphisms\/} $\{g_L^{\Delta x_L}\}$ \resp{$\{g_R^{\Delta x_R}\}$} shifting each point $x_L$ \resp{$x_R$} of the string by a (small) interval (of arc) $\Delta x_L$ \resp{$\Delta x_R$}~.
}
\vskip 11pt

\bpr
\Bi
\item Indeed, the action of the one-parameter group of diffeomorphisms 
 $\{g_L^{x_L}\}$ \resp{$\{g_R^{x_R}\}$} on the \lr string $\phi  ^S_{L_\mu }(x_L)$ \resp{$\phi  ^S_{R_\mu }(x_R)$} is a homomorphism:
 \[ \{g_L^{x_L}\}\To \Diff(\phi  ^S_{L_\mu }(x_L))\qquad
 \rresp{\{g_R^{x_R}\}\To \Diff(\phi  ^S_{R_\mu }(x_R))}\]
 such that the induced map
 \begin{align*}
 \{g_L^{x_L}\} \times \phi  ^S_{L_\mu }(x_L) &\To \phi  ^S_{L_\mu }(x_L)\\
 \rresp{\{g_R^{x_R}\} \times \phi  ^S_{R_\mu }(x_R) &\To \phi  ^S_{R_\mu }(x_R)}\end{align*}
 is differentiable \cite{Sma}.
 
 \item If the \lr string $\phi  ^S_{L_\mu }(x_L)$ \resp{$\phi  ^S_{R_\mu }(x_R)$} is a circle as envisaged in section 4.9, then the one-parameter group of diffeomorphisms $\{g_L^{x_L}\}$ \resp{$\{g_R^{x_R}\}$} corresponds to a rotation of $\phi ^S_{L_\mu }(x_L)$ \resp{$\phi ^S_{R_\mu }(x_R)$} by a small (interval of) arc of circle.\epr
 \Ei
 \vskip 11pt
 
\subsection{The structure of the basic diagrams of AQT}

\Bi
\item A basic diagram of AQT is of the type:
\vskip 11pt
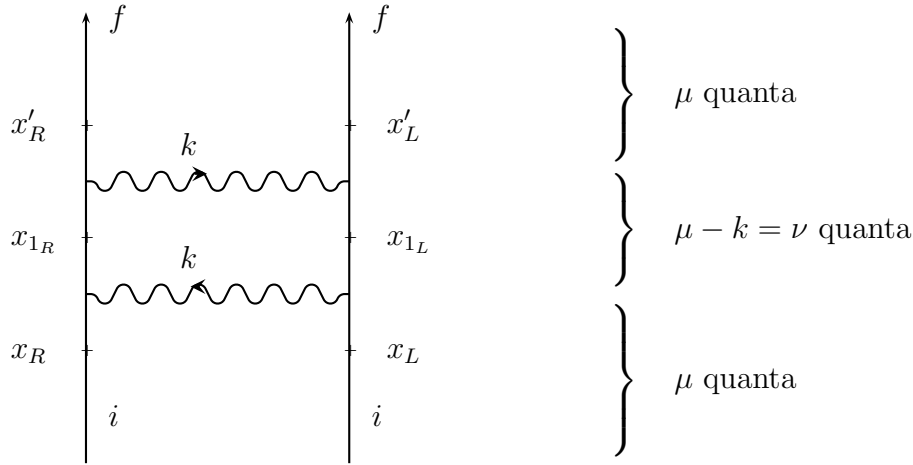
\centerline{\begin{pspicture}(13,6)
\psline{->}(1.5,0)(1.5,6)
\psline{->}(5,0)(5,6)
\pszigzag[coilarm=.05,linearc=.1,coilwidth=.5](1.5,3.75)(5,3.75)
\pszigzag[coilarm=.05,linearc=.1,coilwidth=.5](1.5,2.25)(5,2.25)
\psdots[dotstyle=+](1.52,1.5)(1.52,3)(1.52,4.5)(5.02,1.5)(5.02,3)(5.02,4.5)
\psline[arrowlength=1,arrowsize=.2]{->}(2.95,3.85)(3.11,3.85)
\psline[arrowlength=1,arrowsize=.2]{<-}(2.89,2.35)(2.95,2.35)
\put(0.5,1.4){$x_R$}
\put(0.5,2.9){$x_{1_R}$}
\put(0.5,4.4){$x'_{R}$}
\put(5.5,1.4){$x_L$}
\put(5.5,2.9){$x_{1_L}$}
\put(5.5,4.4){$x'_{L}$}
\put(1.8,0.5){$i$}
\put(1.8,5.8){$f$}
\put(5.3,0.5){$i$}
\put(5.3,5.8){$f$}
\put(2.75,2.6){$k$}
\put(2.75,4.1){$k$}
\put(8,1){$\L.\bt{c}\mbox{}\\[6pt] \mbox{}\te\R\} \quad \mu $ quanta}
\put(8,4.8){$\L.\bt{c}\mbox{}\\ \mbox{}\te\R\} \quad \mu $ quanta}
\put(8,3){$\L.\bt{c}\mbox{}\\[-6pt] \mbox{}\te\R\} \quad \mu-k=\nu  $ quanta}
\end{pspicture}}
\vskip 11pt

where:
\Bi
\item  the right and left vertical lines represent the rotations respectively of the current points $x_R$ and $x_L$ of the right and left strings $\phi ^S_{R_\mu }(x_R)$ and $\phi ^S_{L_\mu }(x_L)$ at $\mu $ quanta from an initial state \mfg{i} to a final state \mfg{f};

\item the lower and upper horizontal lines represent respectively the exchange from ``left to right'' and from ``right to left'' of $k$ magnetic, electric or gravitational quanta during a small interval of time $\Delta t_1$ corresponding to a rotation of arcs $\Delta x_{1_R}$ and $\Delta x_{1_L}$~.
\Ei

\item As $\phi ^S_{R_\mu }(x_R)$ rotates in the opposite sense with respect to $\phi ^S_{L_\mu }(x_L)$~, we recover the evolutionary interpretation of particles and antiparticles of the diagrams of Feynman.

\item The exchanges of $k$ biquanta between the left and right strings result from the action of a (bi)potential on these as it will be shown in the following.

\item As it will be seen, {\bbf there are two kinds of left and right propagators of Green\/}:
\Bena
\item The firsts $G^{(\mu )}_{0_L}(x'_L;x_L)$ and $G^{(\mu )}_{0_R}(x'_R;x_R)$  rotate respectively the left and right points $x_L$ and $x_R$ of the strings $\phi ^S_{L_\mu }(x_L)$ and $\phi ^S_{R_\mu }(x_R)$ at $\mu $ quanta by the arcs of circle $(x'_L-x_L)$ and $(x'_R-x_R)$~.

\item the seconds $G^{(\mu-k )}_{1_L}(\Delta x_{1_L})$ and $G^{(\mu-k )}_{1_R}(\Delta x_{1_R})$ rotate respectively the same left and right strings acted by a bipotential provoking exchanges of $k$ biquanta between them under their rotations of the arcs of circles $\Delta x_{1_L}$ and $\Delta x_{1_R}$ during an interval of time $\Delta t_1$~.

As a consequence, the strings $\phi ^S_{L_\mu }(x_L)$ and $\phi ^S_{R_\mu }(x_R)$ are transformed during $\Delta t_1$ into strings $\phi ^S_{L_{(\mu -k)}}(x_{1_L})$ and $\phi ^S_{R_{(\mu -k)}}(x_{1_R})$ at $(\mu -k)$ quanta, the remaining left and right $k$ quanta travelling between them.
\Ee
\Ei
\vskip 11pt

\subsection{Free rotating bistrings}

\Bi
\item We shall now study more precisely the Green's propagators and the $S$-matrix elements of left and right ``space'' strings (at the \mfg{ST},  \mfg{MG} and  \mfg{M} levels) interacting by means of magnetic subfields by the exchanges of magnetic biquanta.

\item Let $(\phi ^S_{R_\mu }(x_R)\otimes_D \phi ^S_{L_\mu }(x_L))$ be such bistring rotating freely, i.e. without the exchange of magnetic biquanta.

Assume that the left and right strings $\phi ^S_{L_\mu }(x_L)$ and $\phi ^S_{R_\mu }(x_R)$ rotate respectively on the arcs of circles $(x'_L-x_L)$ and $(x'_R-x_R)$ in such a way that {\bbf the corresponding evolution\/} in time during $\Delta t=(t'-t)$ {\bbf of the bistring be given by\/}:

\begin{multline*}
\phi ^S_{R_\mu }(x'_R;t') \otimes_D 
\phi ^S_{L_\mu }(x'_L;t') \\
= (-i\int dx_R\ G^{(\mu )}_{0_R}(x'_R;x_R)\ \phi ^S_{R_\mu }(x_R;t))
\otimes_D (+i\int dx_L\ G^{(\mu )}_{0_L}(x'_L;x_L)\ \phi ^S_{L_\mu }(x_L;t))
\end{multline*}
where $G^{(\mu )}_{0_L}(x'_L;x_L)$ \resp{$G^{(\mu )}_{0_R}(x'_R;x_R)$} is the free \lr propagator of Green at $\mu $ quanta.
\Ei
\vskip 11pt

\subsection{The context of the exchange of magnetic biquanta into a bistring}

\Bi
\item If the bistring $(\phi ^S_{R_\mu }(x_R)\otimes_D \phi ^S_{L_\mu }(x_L))$ does no more rotate freely, then, it can be submitted to {\bbf the magnetic biendomorphism $(E_{R_\mu }\otimes_m E_{L_\mu })$} considered in section 2.4 and transforming it into:
\[ (\phi ^S_{R_\mu }(x_{1_R})\otimes \phi ^S_{L_\mu }(x_{1_L}))
= (\phi ^S_{R_\nu }(x_{1_R})\otimes_D \phi ^S_{L_\nu }(x_{1_L}))
+ \sum^k_{\rho =1}(\widetilde M^S_{R_\rho }\otimes_m\widetilde M^S_{L_\rho })\]
where $(\widetilde M^S_{R_\rho }\otimes_m\widetilde M^S_{L_\rho })$ is one of the magnetic biquanta (functions) exchanged inside the bistring $(\phi ^S_{R_\mu }(x_{1_R})\otimes \phi ^S_{L_\mu }(x_{1_L}))$~.

Consequently, the exchange of the $k$ magnetic biquanta inside
$(\phi ^S_{R_\mu }(x_{1_R})\otimes \phi ^S_{L_\mu }(x_{1_L}))$ during a small interval of time $\Delta t_1$ transforms it into the ``diagonal'' bistring\linebreak
$(\phi ^S_{R_\nu }(x_{1_R})\otimes_D \phi ^S_{L_\nu }(x_{1_L}))$ at $\nu $ biquanta during the rotations of the arcs of circle $\Delta x_{1_R}$ and $\Delta x_{1_L}$~.

\item On the other hand, if we refer to the differential wave equation of the bisemiphoton in a three-dimensional Cartesian coordinate frame as envisaged in chapter 4 of \cite{Pie2}, we see that a left and right unitary inner automorphisms transform this equation into the $1D$-degenerate second order differential equation of the harmonic oscillator which is that of the bistring
$(\phi ^S_{R_\mu }(x_{R})\otimes_D \phi ^S_{L_\mu }(x_{L}))$~.

It then results that {\bbf there is a one-to-one correspondence between a bisemiphoton at $\mu $ biquanta, in a $3D$-Cartesian frame, submitted to the magnetic (bi)potential\/}
\[ \underset{i\neq j}{\sum^3_{i=1}\sum^3_{j=1}} \not{p}_i\ \not{p}_j=
\underset{i\neq j}{\sum_{i}\sum_{j}} \F{\hbar^2}{c^2}\ s^i\ s^j\ \F\partial{\partial x_i}\ \F\partial{\partial x_j}\]
{\bbf and the bistring at $\mu $ biquanta\/} $(\phi ^S_{R_\mu }(x_{R})\otimes \phi ^S_{L_\mu }(x_{L}))$ {\bbf submitted to the magnetic biendomorphism $(E_{R_\mu }\otimes E_{L_\mu })$}.

So, the exchange of $k$ magnetic biquanta is generated by the magnetic potential corresponding mathematically to the magnetic biendomorphism.

\Ei
\vskip 11pt

\subsection{Solutions in  terms of (bi)propagators in the magnetic case}

\Bi
\item More concretely, the bilinear  {\bbf differential equations of the bistring\/} 
$(\phi ^S_{R_\mu }(x_{R})\otimes \phi ^S_{L_\mu }(x_{L}))$ is:
\[ (\phi ^{S_p}_{R_\mu }(x_{1_R})\otimes \phi ^{S_p}_{L_\mu }(x_{1_L}))
= \lambda ^2_\mu (\phi ^S_{R_\mu }(x_{1_R})\otimes \phi ^S_{L_\mu }(x_{1_L}))\]
or
\[ \L(
\F\hbar c\ s_{x_{1_{R_\mu }}}\ \F{\partial \phi ^{S}_{R_\mu }(x_{1_R})}{\partial x_{1_R}}
\otimes
\F\hbar c\ s_{x_{1_{L_\mu }}}\ \F{\partial \phi ^{S}_{L_\mu }(x_{1_L})}{\partial x_{1_L}}
\R)
= \lambda ^2_\mu (\phi ^S_{R_\mu }(x_{1_R})\otimes \phi ^S_{L_\mu }(x_{1_L}))\]
in such a way that the bigenerator of the considered Lie bialgebra is given by the eigen(vi)value 
$ \lambda ^2_\mu = \lambda ^2_{D_{(\mu -k)}}+\lambda ^2_{m_{\mu _k}}$ which splits into a diagonal part $\lambda ^2_{D_{(\mu -k)}}$ at $(\mu -k)$ biquanta and into a magnetic part $\lambda ^2_{m_{\mu _k}}$ at $k$ biquanta.

\item {\bf Its solution\/} is given by:
\begin{multline*}
(\phi ^S_{R_\mu }(x'_R;t')
\otimes \phi ^S_{L_\mu }(x'_L;t'))\\
\begin{aligned}
= &\L[\L(-i\int dx_{1_R}\ G^{(\mu -k)}_{1_R}(x'_R;x_{1_R})\ \phi ^S_{R_{(\mu -k)}}(x_{1_R})\R)\R.\\
&\quad \times 
\L.\L(+i\int dx_{1_L}\ G^{(\mu -k)}_{1_L}(x'_L;x_{1_L})\ \phi ^S_{L_{(\mu -k)}}(x_{1_L})\R)\R]\\
& \qquad + \L[\L(-i\iint dx_{1_R}\ dx_{1_L}\ G^{k}_{1_{L\to R}}(x_{1_R};x_{1_L})\ \L(\sum^k_{\rho =1}\widetilde M^S_{L_\rho }(x_{1_R};x_{1_L})\R)\R)\R.\end{aligned}\\
 \times \L.\L(+i\iint dx_{1_L}\ dx_{1_R}\ G^{k}_{1_{R\to L}}(x_{1_L};x_{1_R})\ \L(\sum^k_{\rho =1}\widetilde M^S_{L_\rho }(x_{1_L};x_{1_R})\R)\R)\R]\end{multline*}
where:
\Bi
\item $G^{(\mu -k)}_{1_L}(x'_L;x_{1_L})$ \resp{$G^{(\mu -k)}_{1_R}(x'_R;x_{1_R})$} is the \lr propagator rotating the \lr string
$\phi ^S_{L_{(\mu -k)}}(x_{1_L})$ \resp{$\phi ^S_{R_{(\mu -k)}}(x_{1_R})$} at $(\mu -k)$ quanta along the arc of circle 
$\Delta x_{1_L}$ \resp{$\Delta x_{1_R}$} during the interval of time $\Delta t_1$ (see section 4.1).

\item $G^{k}_{1_{L\to R}}(x_{1_R};x_{1_L})$
\resp{$G^{k}_{1_{R\to L}}(x_{1_L};x_{1_R})$} is the  propagator responsible for the exchange of $k$ quanta, $1\le k\le \mu $~, from the \lr string
$\phi ^{S}_{L_{(\mu -k)}}(x_{1_L})$
\resp{$\phi ^{S}_{R_{(\mu -k)}}(x_{1_R})$} to the \rl string
$\phi ^{S}_{R_{(\mu -k)}}(x_{1_R})$ \resp{$\phi ^{S}_{L_{(\mu -k)}}(x_{1_L})$}
during the rotations of these strings of the arcs of circles $\Delta x_{1_L}$ and $\Delta x_{1_R}$~.
\Ei

According to proposition 4.10, these propagators are one-parameter groups of diffeomorphisms.

\item The corresponding {\bbf scattering matrix intensity\/} will be given by:
\begin{multline*}
 {}_{fi}S_{R_\mu }
\times {}_{fi}S_{L_\mu }\\
= \lim_{t'\to\infty }\int dx'_R\ {}_f\phi ^S_{R_\mu }(x'_R;t')\ {}_i\psi^S_{R_\mu }(x'_R;t')
\times \lim_{t'\to\infty }\int dx'_L\ {}_f\phi ^S_{L_\mu }(x'_L;t')\ {}_i\psi^S_{L_\mu }(x'_L;t')
\end{multline*}
where:
\Bi
\item the indices \mfg{i} and \mfg{f} refer respectively to initial and final states;
\item ${}_f\phi ^S_{R_\mu }(x'_R;t')\equiv \phi ^S_{R_\mu }(x'_R;t')$ such that
$\phi ^S_{R_\mu }(x'_R;t')$ is the above mentioned solution when $t'\to +\infty $~;
\item ${}_i\psi ^S_{R_\mu }(x'_R;t')= \phi ^{*S}_{R_\mu }(x'_R;t')$ such that
$\phi ^{*S}_{R_\mu }(x'_R;t')$ is the complex conjugate solution when $t'\to-\infty $~.
\Ei

Remark that this scattering matrix intensity is unitary if the solutions are normalized, i.e. if
$\phi ^S_{L_\mu }(x'_L;t')$ and $\phi ^S_{R_\mu }(x'_R;t')$ are divided respectively by their amplitudes $c_\mu $ and $c^*_\mu $ according to section 4.9.
\Ei
\vskip 11pt

\subsection{Exchange of biquanta in an electric bistring}

\Bi
\item The electric case at the \mfg{ST}, \mfg{MG} or \mfg{M} level can be handled as it was done for the magnetic case.  But, the bistring to be considered is then of the type
\[ \phi ^T_{R_\mu }(t_R)\otimes \phi ^S_{L_\kappa  }(x_L)\qquad
\text{(or} \quad 
 \phi ^S_{R_\kappa  }(x_R)\otimes \phi ^T_{L _\mu }(t_L)\ )\]
 i.e. given by the product of a right ``time'' string at $\mu $ quanta by a left ``space'' string at $\kappa $ quanta (or by the product of a right ``space'' string at $\kappa$ quanta by a left ``time'' string at $\mu $ quanta).
 
 \item This bistring is rotating in the corresponding Lie bialgebra and is given there by the operator valued bistring $\phi ^{T_p}_{R_\mu }(t_R)\otimes \phi ^S_{L_\kappa }(x_L)$ which is in bijection with its equivalent considered in a $3D$-Cartesian coordinate frame and submitted to the electric (bi)potential 
 $\sum\limits^3_{i=1}\not{m}_0\ \not{p}_i$ as it was developed in proposition 2.8.
 
\item {\bf The bistring $\phi ^T_{R_\mu }(t_R)\otimes \phi ^S_{L_\mu }(x_L)$~, submitted to an electric (bi)potential resulting essentially from the difference of rotational velocities of 
$\phi ^T_{R_\mu }(t_R)$ and of $\phi ^S_{L_\kappa  }(x_L)$~, emits a set of \mfg{\ell} electric biquanta of exchange by means of the electric biendormorphism\/}
\begin{multline*}
E_{R_\mu }\otimes_e E_{L_\kappa }: \quad
(\phi ^T_{R_\mu }(t_{1_R})\otimes \phi ^S_{L_\kappa  }(x_{1_L}))\\
\To (\phi ^T_{R_\nu }(t_{1_R})\otimes_{D_e} \phi ^S_{L_\lambda  }(x_{1_L}))
+ \L(\sum^\ell_{\rho =1}(\widetilde M^T_{R_\rho }\otimes_e \widetilde M^S_{L_\rho })\R)
\end{multline*}
where $\mu -\nu =\kappa -\lambda =\ell $~.

So, the exchange of $\ell $ electric biquanta inside
$(\phi ^T_{R_\mu }(t_{1_R})\otimes \phi ^S_{L_\kappa  }(x_{1_L}))$ during a small interval of time $\Delta t_1$ transforms it into the diagonal electric bistring
$(\phi ^T_{R_\nu }(t_{1_R})\otimes_{D_e} \phi ^S_{L_\lambda  }(x_{1_L}))$ at $\nu $ and $\lambda $ quanta during the rotations of the arcs of circle 
$\Delta t_{1_R}$ and
$\Delta t_{1_L}$~.

\item The corresponding bilinear differential equation is:
\[ \L(\F\hbar c\ s_{0_{R_\mu }}\ \F{\partial \phi ^T_{R_\mu }(t_{1_R})}{\partial t_{1_R}}\R)
\otimes \L(\F\hbar c\ s_{x_{1_L}}\ \F{\partial \phi ^S_{L_\kappa  }(x_{1_L})}{\partial x_{1_L}}\R)
= \lambda _\mu \centerdot \lambda _\kappa \ \L(\phi ^T_{R_\mu }(t_{1_R})
\otimes \phi ^S_{L_\kappa  }(x_{1_L})\R)\]
where the generator of the considered Lie bialgebra $\lambda _\mu \centerdot \lambda _\kappa= \lambda ^2_{D_{(\nu ,\lambda )}}+\lambda ^2_{e_\ell }$ splits into a diagonal part $\lambda ^2_{D_{(\nu ,\lambda )}}$ at $\nu $ and $\lambda $ quanta and into an electric part $\lambda ^2_{e_\ell }$ at $\ell $ biquanta.

It solution is then given by:
\begin{multline*}
(\phi ^T_{R_\mu }(t'_R;t')
\otimes \phi ^S_{L_\kappa  }(x'_L;t'))\\
\begin{aligned}
= &\L[\L(-i\int dt_{1_R}\ G^{(\mu -\ell )}_{1_R}(t'_R;t_{1_R})\ \phi ^T_{R_{(\mu -\ell )}}(t_{1_R})\R)\R.\\
&\quad \times 
\L.\L(+i\int dx_{1_L}\ G^{(\kappa  -\ell )}_{1_L}(x'_L;x_{1_L})\ \phi ^S_{L_{(\kappa  -\ell )}}(x_{1_L})\R)\R]\\
& \qquad + \L[\L(-i\iint dt_{1_R}\ dx_{1_L}\ G^{\ell }_{1_{L\to R}}(t_{1_R};x_{1_L})\ \L(\sum^\ell _{\rho =1}\widetilde M^S_{L_\rho }(t_{1_R};x_{1_L})\R)\R)\R.\end{aligned}\\
 \times \L.\L(+i\iint dx_{1_L}\ dt_{1_R}\ G^{\ell }_{1_{R\to L}}(x_{1_L};t_{1_R})\ \L(\sum^\ell _{\rho =1}\widetilde M^T_{R_\rho }(x_{1_L};t_{1_R})\R)\R)\R]\end{multline*}
and illustrated by the diagram:
\vskip 11pt
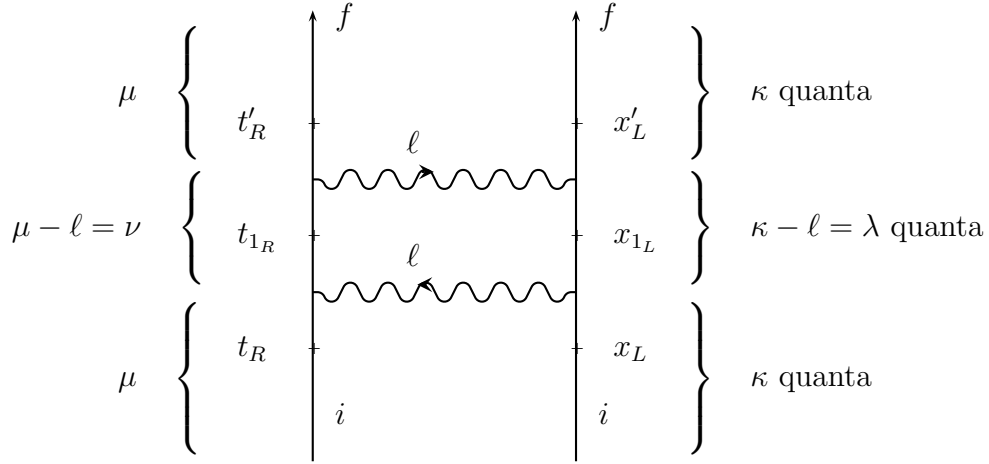
\centerline{\hspace*{4cm}\begin{pspicture}(11,6)
\psline{->}(1.5,0)(1.5,6)
\psline{->}(5,0)(5,6)
\pszigzag[coilarm=.05,linearc=.1,coilwidth=.5](1.5,3.75)(5,3.75)
\pszigzag[coilarm=.05,linearc=.1,coilwidth=.5](1.5,2.25)(5,2.25)
\psdots[dotstyle=+](1.52,1.5)(1.52,3)(1.52,4.5)(5.02,1.5)(5.02,3)(5.02,4.5)
\psline[arrowlength=1,arrowsize=.2]{->}(2.95,3.85)(3.11,3.85)
\psline[arrowlength=1,arrowsize=.2]{<-}(2.89,2.35)(2.95,2.35)
\put(0.5,1.4){$t_R$}
\put(0.5,2.9){$t_{1_R}$}
\put(0.5,4.4){$t'_{R}$}
\put(5.5,1.4){$x_L$}
\put(5.5,2.9){$x_{1_L}$}
\put(5.5,4.4){$x'_{L}$}
\put(1.8,0.5){$i$}
\put(1.8,5.8){$f$}
\put(5.3,0.5){$i$}
\put(5.3,5.8){$f$}
\put(2.75,2.6){$\ell $}
\put(2.75,4.1){$\ell $}
\put(6,1){$\L.\bt{c}\mbox{}\\[6pt] \mbox{}\te\R\} \quad \kappa  $ quanta}
\put(6,4.8){$\L.\bt{c}\mbox{}\\ \mbox{}\te\R\} \quad \kappa  $ quanta}
\put(6,3){$\L.\bt{c}\mbox{}\\[-6pt] \mbox{}\te\R\} \quad \kappa -\ell =\lambda   $ quanta}
\put(-1.5,1){$\quad \mu \quad \L\{\bt{c}\mbox{}\\[6pt] \mbox{}\te\R.     $}
\put(-1.5,4.8){$\quad \mu \quad \L\{\bt{c}\mbox{}\\ \mbox{}\te\R.    $}
\put(-2.5,3){$\mu  -\ell =\nu \quad\L\{\bt{c}\mbox{}\\[-6pt] \mbox{}\te\R.      $}
\end{pspicture}}
\vskip 11pt

in such a way that:
\Bi
\item $G^{(\kappa  -\ell )}_{1_L}(x'_L;x_{1_L})$ \resp{$G^{(\mu   -\ell )}_{1_R}(t'_R;t_{1_R})$} is the \lr propagator rotating the \lr string
$\phi ^S_{L_{(\kappa  -\ell )}}(x_{1_L})$
\resp{$\phi ^T_{R_{(\mu   -\ell )}}(t_{1_R})$} at $(\kappa -\ell )$ \resp{$(\mu -\ell )$} quanta along the arc of circle $\Delta x_{1_L}$ \resp{$\Delta t_{1_R}$} during the interval of time $\Delta t_1$~.

\item $G^\ell_{1_{L\to R}}(t_{1_R},x_{1_L})$ \resp{$G^\ell_{1_{R\to L}}(x_{1_L},t_{1_R})$} is the propagator responsible for the exchange of $\ell $ quanta from the \lr space \resp{time} string
$\phi ^S_{L_{(\kappa  -\ell )}}(x_{1_L})$
\resp{$\phi ^T_{R_{(\mu   -\ell )}}(t_{1_R})$} to the \rl string
$\phi ^T_{R_{(\mu   -\ell )}}(t_{1_R})$
\resp{$\phi ^S_{L_{(\kappa  -\ell )}}(x_{1_L})$} during the rotations of these strings of the arcs of circle $\Delta x_{1_L}$ and $\Delta t_{1_R}$~.
\Ei
\Ei
\vskip 11pt

\subsection{Remarks of the Feynman paths in AQT and QFT}

\Bena
\item The other diagrams considered in section 4.8 can be handled similarly. They are of the same type and describe the exchanges of gravitational, electric and magnetic biquanta.

\item If we refer to the original paper \cite{Fey1} of R.P. Feynman, we see that {\bbf his different paths\/} from a point $A$ to a point $B$ {\bbf correspond in AQT to the possible normal modes of the exchanged biquanta\/}.  For example, in section 4.13, every path of Feynman would correspond to a value of \mfg{\rho } of the number of exchanged biquanta, \ldots\ between a right a a left string.
\Ee
\vskip 11pt

\subsection{Proposition}
{\em {\bfseries The algebraic quantum theory\/} (AQT), which is a mathematical theory of the Physics of elementary particles, {\bfseries is free of divergences\/}.}
\vskip 11pt

\bpr This results essentially from the facts that:
\Bena
\item {\bf the interactions} between fields (and strings) {\bf are not realized perturbatively in AQT} according to proposition 4.4.

Consequently, infinite perturbative series have not to be considered.

\item The basic diagrams in AQT concern the exchanges of gravitational, magnetic and electric biquanta between left and right strings and correspond in QFT to first order perturbative diagrams implying the exchanges of virtual photons between fermions.\epr
\Ee
\vskip 11pt

\subsection{Corollary}

{\em {\bfseries The famous self energy diagram in QFT} corresponds in AQT to a basic diagram between a right and a left string belonging to the space ``mass'' (\mfg{M}) field of an elementary bisemifermion and exchanging a magnetic biquantum.}
\vskip 11pt

\centerline{\bt{ccc}
\begin{pspicture}(1,2.5)
\psline{->}(0,0)(0,2.5)
\psline(0,0.5)(0.5,0.5)
\psline(0,2)(0.5,2)
\pszigzag[coilarm=.05,linearc=.1,coilwidth=.5](0.5,0.5)(0.5,2)
\end{pspicture}
& \raisebox{1cm}{$\xrightarrow{\quad \sim  \quad}$} &
\begin{pspicture}(3.5,2.5)
\psline{->}(0,0)(0,2.5)
\psline{->}(3.5,0)(3.5,2.5)
\pszigzag[coilarm=.05,linearc=.1,coilwidth=.5](0,1.75)(3.5,1.75)
\pszigzag[coilarm=.05,linearc=.1,coilwidth=.5](0,0.75)(3.5,0.75)
\psline[arrowlength=1,arrowsize=.2]{->}(1.45,1.85)(1.61,1.85)
\psline[arrowlength=1,arrowsize=.2]{<-}(1.39,0.85)(1.45,0.85)
\end{pspicture}
\\
\bt[t]{c}
self-energy\\[-6pt] diagram in QFT\te
&&
\bt[t]{c}
diagram of the exchange\\[-6pt]
 of one magnetic biquantum\\[-6pt]
 inside a bistring.\te
 \te}

\addtocontents{toc}{\eject}

\eject

\end{document}